\documentclass[5p,authoryear,times,twocolumn]{elsarticle} 

\usepackage{natbib}
\usepackage{color}
\usepackage[latin1]{inputenc}
\usepackage{graphicx}
\usepackage{rotating}
\usepackage{booktabs}
\usepackage{amsmath}
\usepackage{amssymb}
\usepackage{lscape}
\usepackage{overpic}
\usepackage{scrextend}
\usepackage{booktabs}
\usepackage{lineno}
\journal{Icarus}
\begin{document}
\begin{frontmatter}
\title{A new method to determine the grain size of planetary regolith}
\author[igep]{B. Gundlach}
\ead{b.gundlach@tu-bs.de}
\author[igep]{J. Blum}
\address[igep]{Institut für Geophysik und extraterrestrische Physik, Technische Universität Braunschweig, \\Mendelssohnstr. 3, D-38106 Braunschweig, Germany}
\begin{abstract}
Airless planetary bodies are covered by a dusty layer called regolith. The grain size of the regolith determines the temperature and the mechanical strength of the surface layers. Thus, knowledge of the grain size of planetary regolith helps to prepare future landing and/or sample-return missions. In this work, we present a method to determine the grain size of planetary regolith by using remote measurements of the thermal inertia. We found that small bodies in the Solar System (diameter less than $\sim 100 \, \mathrm{km}$) are covered by relatively coarse regolith grains with typical particle sizes in the millimeter to centimeter regime, whereas large objects possess very fine regolith with grain sizes between 10 $\rm \mu m$ and 100 $\rm \mu m$.
\end{abstract}
\end{frontmatter}

\setlength{\tabcolsep}{10pt}
\renewcommand{\arraystretch}{2}
\renewcommand{\topfraction}{1}
\renewcommand{\bottomfraction}{1}
\section{Introduction}
Planetary surfaces are exposed to a continuous flux of impactors of various sizes, which have, due to the hyper-velocity nature of the impacts, ground down the initially rocky material to ever finer particle sizes \citep{Chapman1976,Housen1982}. Thus, airless bodies in the Solar System (e.g., Mercury, the Moon, planetary moons, and asteroids) are covered by planetary regolith, a granular material consisting of distinct solid grains \citep{Chapman2004}. Each hyper-velocity impact causes the ejection of material from the forming crater, which can partly be re-captured by the planetary body if its gravitational escape speed exceeds the ejection velocity \citep{Housen1982,Cintala1978}. Fragments with higher velocities are lost into interplanetary space.
\par
Laboratory experiments \citep{Fujiwara1980,Nakamura1991,Nakamura1993,Nakamura1994} and studies of the crater structures on the Moon \citep{Vickery1986,Vickery1987} have shown that hyper-velocity impacts accelerate small fragments to higher velocities than large ones so that we expect that larger planetary bodies possess a finer average regolith grain size than small bodies \citep{Cintala1978}. However, the mean particle size of planetary regolith has, except for the Moon \citep{Duke1970,McKay2009}, not been measured directly.
\par
In this work, we present a new method to determine the grain size of planetary regolith from remote or in-situ thermal inertia measurements (see Sect. \ref{Strategy} and Fig. \ref{Fig_flow_chart}). Using literature data of the heat conductivity of lunar regolith \citep{Cremers1971a,Cremers1971b}, sampled by the Apollo 11 and 12 astronauts, we fit our published heat-conductivity model of granular material \citep[][see Sect. \ref{The heat conductivity model}]{Gundlach2012} to the lunar regolith and show that the resulting characteristic grain size agrees with the regolith size distribution measured for the Moon \citep[][see Sect. \ref{Calibration of the heat conductivity model using lunar regolith}]{McKay2009}. We confirm the expected anti-correlation between the regolith grain size and the gravitational acceleration of the planetary body for a large number of asteroids, the Moon, the Martian moons and Mercury with diameters between 0.3 km and 4,880 km (see Sects. \ref{Important properties of the analyzed objects}-\ref{Influence of the network heat conductivity on the grain size estimation}). Our results can help to prepare future landing and sample-return missions to primitive bodies of the Solar System (see Sect. \ref{Discussion}).

\section{Strategy}\label{Strategy}
\begin{figure}[t]
\centering
\includegraphics[angle=0,width=1.0\columnwidth]{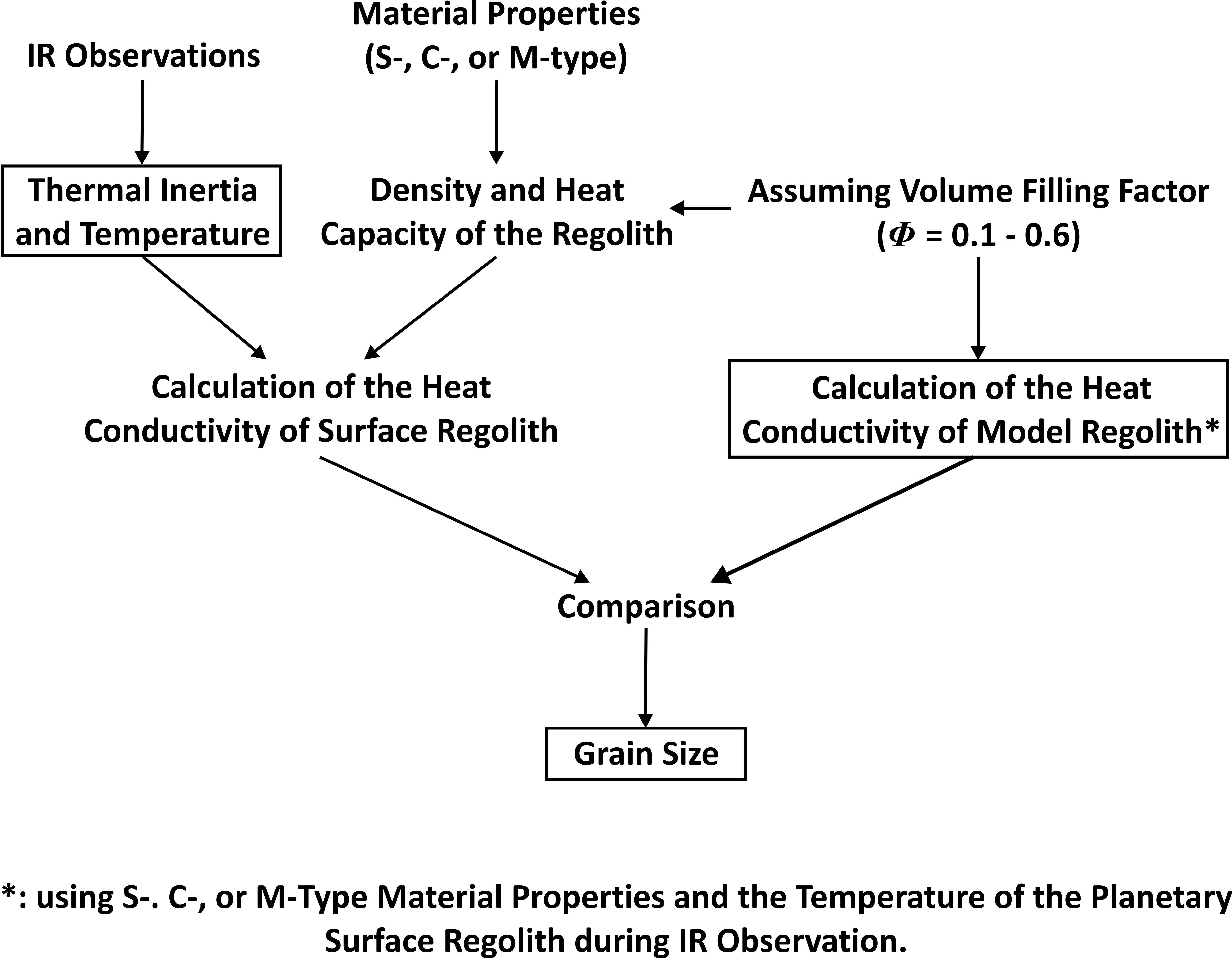}
\caption{Strategy for the grain size determination of planetary regolith using thermal-inertia measurements.}
\label{Fig_flow_chart}
\end{figure}

To determine the particle size of planetary regolith (see Fig. \ref{Fig_flow_chart} for a flow chart of our approach), we used literature data of thermal inertiae measured for different airless bodies in the Solar System (see Sect. \ref{Important properties of the analyzed objects}). The thermal inertia,
\begin{align}
\Gamma \, = \, \sqrt{\lambda \, C}  \, \mathrm{,}
\label{eq0}
\end{align}
describes the resistance of the near surface material of a Solar System body to follow diurnal changes in the irradiation and depends on the heat conductivity $\lambda$ and the volumetric heat capacity of the bulk regolith, respectively. The volumetric heat capacity can be expressed by $C \, = \, \phi \, \rho \, c$, with the packing fraction $\phi$, the mass density $\rho$, and the specific heat capacity $c$ of the regolith particles. Thus, the thermal inertia depends on the material properties of the regolith and the degree of compaction of the regolith particle layers. We also expect that the heat conductivity depends on the size of the regolith particles as suggested by \citet{Gundlach2012}. With the knowledge of the thermal inertia, the surface-material properties and assumptions about the packing density of the regolith particles, one can thus derive the particle size of the surface regolith.
\par
For the estimation of the mass density and the specific heat capacity of the regolith material, we divided the analyzed bodies into stony (S), carbonaceous (C), and metallic (M) objects (see Sect. \ref{Important properties of the analyzed objects}). For each of the three classes, the mass density and heat capacity of the material were approximated by the laboratory measurements of these properties of representative meteorites \citep{Opeil2010}, i.e. Cronstad and Lumpkin (S), Cold Bokkeveld and NWA 5515 (C) and Campo del Cielo (M) (see Sect. \ref{The heat conductivity model}). In order to take different degrees of compaction of the regolith into account, we treated the volume filling factor (or packing fraction) of the material as a free parameter and varied it between $\phi \,= \,0.1$ and $\phi \, = \,0.6$ in intervals of $\Delta \phi \, = \, 0.1$. For each of the six packing densities, we thus arrive at a heat conductivity value. In general, the heat capacity and the thermal conductivity are temperature dependent. Thus, we also derived the surface temperatures of the celestial bodies in our sample at the time of observation of their thermal inertiae. The mean grain size of the planetary regolith can then be determined from a comparison with a modeled heat conductivity of granular materials in vacuum \citep[][see Sect. \ref{Determination of the regolith particle size}]{Gundlach2012}.

\section{The heat conductivity model}\label{The heat conductivity model}
\begin{table*}[p!]
\begin{center}
    \scriptsize
    \caption{Summary of the material properties used in the model.}\vspace{1mm}
    \begin{tabular}{lccccc}
        \bottomrule
        Physical Property & Value & Unit & Ref. & Material \\
        \midrule
        $\mu$                       &  $0.25 $                                                           & -                                  &\citep{Schultz1995}        & Basalt                           \\
        $E$                         &  $(7.8 \, \pm \, 1.9) \times 10^{10}$                              &$\mathrm{[Pa]}$                     &\citep{Schultz1995}        & Basalt                           \\
        $\gamma(T)$                 &  $6.67\times10^{-5} \, T[\mathrm{K}]$                              &$\mathrm{[J \, m^{-2}]}$            &\citep{Gundlach2012}       & $\mathrm{SiO_2}$                 \\
        $\lambda_{\rm solid,S}$     &  $2.18$                                                            &$ \mathrm{[W \, m^{-1} \, K^{-1}]}$ &\citep{Opeil2010}          & Cronstad and Lumpkin (Meteorites) \\
        $\lambda_{\rm solid,C}(T)$  &  $1.19 \, + \, 2.1 \times10^{-3} \, T[\mathrm{K}]          $        &$ \mathrm{[W \, m^{-1} \, K^{-1}]}$ &\citep{Opeil2010}          & Cold Bokkeveld and NWA 5515 (Meteorites)       \\
        $\lambda_{\rm solid,M}(T)$  &  $12.63 \, + \, 5.1 \times10^{-2} \, T[\mathrm{K}]           $                 &$ \mathrm{[W \, m^{-1} \, K^{-1}]}$ &\citep{Opeil2010}          & Campo del Cielo (Meteorite)      \\
        $\rho_{\rm S}$              &  $3700$                                                            &$ \mathrm{[kg \, m^{-3}         ]}$ &\citep{Opeil2010}          & Cronstad and Lumpkin (Meteorites) \\
        $\rho_{\rm C}$              &  $3110$                                                            &$ \mathrm{[kg \, m^{-3}         ]}$ &\citep{Opeil2010}          & Cold Bokkeveld and NWA 5515 (Meteorites)       \\
        $\rho_{\rm M}$              &  $7800$                                                            &$ \mathrm{[kg \, m^{-3}         ]}$ &\citep{Opeil2010}         & Campo del Cielo (Meteorite)      \\
        $c_{\rm S}(T=200\,\mathrm{K})$&  $500$                                                             &$ \mathrm{[J \, kg^{-1} \,K^{-1}]}$ &\citep{Opeil2010}          & Cronstad and Lumpkin (Meteorites) \\
        $c_{\rm C}(T=200\,\mathrm{K})$&  $560$                                                             &$ \mathrm{[J \, kg^{-1} \,K^{-1}]}$ &\citep{Opeil2010}          & Cold Bokkeveld and NWA 5515 (Meteorites)     \\
        $c_{\rm M}(T=200\,\mathrm{K})$&  $375$                                                             &$ \mathrm{[J \, kg^{-1} \,K^{-1}]}$ &\citep{Opeil2010}          & Campo del Cielo (Meteorite)      \\
        $\epsilon_{S,C}$              &  $1.00$                                                               & -                                  & -                                  & -                                \\
        $\epsilon_M$              &  $0.66$                                                               & -                                  & -                                  & -                                \\
        \bottomrule
        \multicolumn{5}{l}{$\mu$: Poisson's ratio.}\\[-0.15cm]
        \multicolumn{5}{l}{$E$: Young's modulus.}\\[-0.15cm]
        \multicolumn{5}{l}{$\gamma(T)$: specific surface energy.}\\[-0.15cm]
        \multicolumn{5}{l}{$\lambda_{\rm solid}(T)$: heat conductivity of the solid material.}\\[-0.15cm]
        \multicolumn{5}{l}{$\rho$: density of the solid material.}\\[-0.15cm]
        \multicolumn{5}{l}{$c$: heat capacity of the solid material.}\\[-0.15cm]
        \multicolumn{5}{l}{$\epsilon$: emissivity of the material.}\\
    \end{tabular}
     \label{Table_2}
     \end{center}
\begin{center}
    \scriptsize
    \caption{Material properties used for the calibration of the heat conductivity model with lunar regolith.}\vspace{1mm}
    \begin{tabular}{lccccc}
        \bottomrule
        Physical Property & Value & Unit & Ref. & Material \\
        \midrule
        $r$                         &  $20 $                                                             & $\mathrm{[\mu m]}$                 &\citep{McKay2009}                 & Lunar Regolith                  \\
        $\mu$                       &  $0.25 $                                                           & -                                  &\citep{Schultz1995}                & Basalt                           \\
        $E$                         &  $(7.8 \, \pm \, 1.9) \times 10^{10}$                              &$\mathrm{[Pa]}$                     &\citep{Schultz1995}                & Basalt                           \\
        $\gamma(T)$                 &  $6.67\times10^{-5} \, T[\mathrm{K}]$                                          &$\mathrm{[J \, m^{-2}]}$            &\citep{Gundlach2012}              & $\mathrm{SiO_2}$      \\
        $\lambda_{\rm solid}$       &  $1.78$                                                            &$ \mathrm{[W \, m^{-1} \, K^{-1}]}$ & -                & average between $\rm SiO_2$ and meteorites\\
        $\rho_{\rm regolith}$       &  $1300$                                                            &$ \mathrm{[kg \, m^{-3}]}$          &\citep{Cremers1971a,Cremers1971b}  & Lunar Regolith                   \\
        $\rho_{\rm solid}$          &  $3040$                                                            &$ \mathrm{[kg \, m^{-3}]}$          &\citep{Warren2001}                  & Lunar Meteorites\\
        $\phi$                      &  $0.43$                                                            & -                                  & -                                          & -                                \\
        $\epsilon$                  &  $1$                                                               & -                                  & -                                  & -                                \\
        \bottomrule
        \multicolumn{5}{l}{$r$: mean radius of the grains.}\\[-0.15cm]
        \multicolumn{5}{l}{$\mu$: Poisson's ratio.}\\[-0.15cm]
        \multicolumn{5}{l}{$E$: Young's modulus.}\\[-0.15cm]
        \multicolumn{5}{l}{$\gamma(T)$: specific surface energy.}\\[-0.15cm]
        \multicolumn{5}{l}{$\lambda_{\rm solid}(T)$: heat conductivity of the solid material.}\\[-0.15cm]
        \multicolumn{5}{l}{$\rho_{\rm regolith}$: density of the regolith.}\\[-0.15cm]
        \multicolumn{5}{l}{$\rho_{\rm solid}$: density of the solid material.}\\[-0.15cm]
        \multicolumn{5}{l}{$\phi$: volume filling factor of the regolith.}\\[-0.15cm]
        \multicolumn{5}{l}{$\epsilon$: emissivity of the material.}\\
    \end{tabular}
     \label{Table_2a}
     \end{center}
\end{table*}

The heat conductivity model for regolith in vacuum used in this work was introduced by \citet{Gundlach2012} and is an extension of the model by \citet{ChanTien1973}. Regolith is a packing of individual solid particles, in which heat can be transported through the solid network of grains and due to radiation inside the pores of the material. Due to the relatively small contact areas between neighboring regolith particles, the heat conductivity through the particle network is generally much lower than the heat conductivity of the solid material the particles consist of. Radiative heat conduction is favored by large void spaces between the regolith particles, i.e. by a large mean free path of the photons.
\par
The heat conductivity of a granular packing of equal-sized spheres with radii $r$ temperature $T$ and packing fraction $\phi$ is given by
\begin{align}
\lambda(r, T, \phi) \, =  \,  \lambda_{\rm solid}(T) \, H(r,T,\phi) \, + \, 8 \, \sigma \, \epsilon \, T^3 \, \Lambda(r,\phi)  \,
\label{eq1}
\end{align}
and consists of a conductive (first term on the rhs of Eq. \ref{eq1}) and a radiative term (second term on the rhs of Eq. \ref{eq1}). It relates the granular heat conductivity to that of the solid material of the regolith particles, $\lambda_{\rm solid}$ through the dimensionless Hertz factor $H(r,T,\phi)$ (describing the reduced heat flux through the contacts between the regolith particles) and the mean free path $\Lambda(r,\phi)$ of the photons within the pore space of the granular material. Here, $\sigma$ and $\epsilon$ denote the Stefan-Boltzmann constant and the material emissivity of the regolith grains, respectively. The mean free path of the photons can be expressed by
\begin{equation}
\Lambda(r,\phi) \, = \, e_{1} \, \frac{1 \, - \, \phi}{\phi} \, r \, \mathrm{,}
\label{eq2}
\end{equation}
with $e_1 \, = \, 1.34 \, \pm \, 0.01$ \citep{Dullien1991, Gundlach2012}. Thus, the radiative heat conductivity is proportional to the regolith grain size.
\par
Due to the expected smallness of the regolith particles and the low gravity environment for asteroids, the inter-particle forces in the uppermost regolith layers are dominated by mutual van der Waals attraction between the regolith grains and not by the particle weight. Thus, the Hertzian dilution factor for such a granular packing can be expressed by \citep{Gundlach2012}
\begin{align}
H(r,T,\phi) \, =  \,   \left[\, \frac{9 \pi}{4} \,\frac{1 \, - \, \mu^2}{E} \, \frac{\gamma(T)}{r} \,\right]^{1/3} \cdot \left(f_1 \, \mathrm{exp}[\, f_2 \, \phi\,]\right) \cdot  \chi \mathrm{.}
\label{eq3}
\end{align}
The first of the three dimensionless factors on the rhs of Eq. \ref{eq3} describes the heat-flux reduction due to the neck between the particles caused by the van der Waals force. Here, $\mu$, $E$, and $\gamma(T)$ are Poisson's ratio, Young's modulus, and the specific surface energy of the grain material, respectively. The derivation of this first dimensionless factor is based on the solution of the heat transfer equation for solid spheres in contact. \citet{ChanTien1973} extended this to a network of monodisperse spheres in contact. The resulting dilution factor depends on the external applied force acting on the particles. This force determines the Hertzian contact area between the particles and, hence the efficiency of the heat conduction through the network of particles. For large particles on Earth ($r \stackrel{>}{\sim} 50 \, \mathrm{\mu m}$), the heat conductivity can be derived using the weight of the particles.
However, for small particles ($r \stackrel{<}{\sim}  50 \, \mathrm{\mu m}$) on Earth, the adhesion force (e.g., van der Waals bonding) between the particles is much larger than their weight. For asteroids, for which the gravitational force is comparatively small, adhesion between the particles is always the dominant force in the first few regolith particle layers, determining the contact area between the regolith particles. In order to implement the adhesive force between the particles into the model, we used the JKR theory \citep{JKR1971} for the calculation of the force between particles in contact. Using the model by \citet{ChanTien1973} and replacing the external force by the internal JKR force, yields the first dimensionless factor on the rhs of Eq. \ref{eq3}.
\par
The second factor on the rhs of Eq. \ref{eq3} contains the structural information about the particle chains inside the regolith, which can transport heat, with $f_1 \, = \, ( \, 5.18 \, \pm \, 3.45 \,) \times 10^{-2}$ and $f_2 \, = \, 5.26 \, \pm \,0.94$ being empirical constants \citep{Gundlach2012}.
\par
Finally, the factor $\chi$ describes the reduction of the heat conductivity when the model assumption of monodisperse spherical particles is relaxed towards irregular polydisperse grains of a real regolith. This factor is determined by the calibration of the heat conductivity model using lunar regolith (see Sect. \ref{Calibration of the heat conductivity model using lunar regolith}).
\par
Using Eqs. \ref{eq1} - \ref{eq3}, the heat conductivity of regolith in vacuum can be expressed as
\begin{align}
\lambda(r, T, \phi) \, =  \, &\lambda_{\rm solid}(T) \cdot \left[\, \frac{9 \pi}{4} \,\frac{1 \, - \, \mu^2}{E} \, \frac{\gamma(T)}{r} \,\right]^{1/3} \nonumber\\
 &\times \left(f_1 \, \mathrm{exp}[\, f_2 \, \phi\,]\right) \cdot  \chi  \,
+ \, 8 \, \sigma \, \epsilon \, T^3 \, e_1 \, \frac{1 \, - \, \phi}{\phi} \, r \, \mathrm{.}
\label{eq6}
\end{align}
Eq. \ref{eq6} shows the dependence of the heat conductivity on the grain size of the regolith, which is the only free parameter of the model if the material properties, the regolith temperature, and the volume filling factor of the regolith are known. For small regolith particle sizes, the first term on the right hand side of Eq. \ref{eq6}  dominates so that $\lambda \,\propto\, r^{-1/3}$. For large regolith particles, high temperatures, and/or very high regolith porosities, the second term dominates, so that we expect $\lambda \,\propto\, r$.
\par
The material properties used to evaluate Eq. \ref{eq6} are summarized in Tab. \ref{Table_2}. For Poisson's ratio and Young's modulus, we used the values measured for basaltic rock. The bulk heat conductivity was estimated from measurements of the heat conductivity of meteorites \citep{Opeil2010}. Therefore, the analyzed objects were divided into three different spectral classes, S (stony objects), C (carbonaceous objects) and M (metallic objects). For each class, representative meteorites were chosen, i.e. Cronstad and Lumpkin  for class S, Cold Bokkeveld and NWA 5515  for class C, and Campo del Cielo  for class M, respectively. Since the meteorite samples are porous, we calculated the bulk heat conductivity of the non-porous material using Maxwell's formula \citep{Maxwell1904},
\begin{align}
\lambda_{\rm solid,non-porous} \, = \, \lambda_{\rm solid,porous} \, \frac{2 \, + \, \psi}{2\, (\, 1 \, - \, \psi \,)} \,  \, \mathrm{.}
\label{eq7}
\end{align}
Here, $\psi = 1- \phi$ is the porosity of the respective meteorite \citep[$\psi = 0.167$ for Cronstad (S), $\psi = 0.191$ for Lumpkin (S), $\psi = 0.373$ for Cold Bokkeveld (C), $\psi = 0.251$ for NWA 5515 (C) and $\psi = 0.012$ for Campo del Cielo (M),][]{Opeil2010}. Maxwell's formula was derived for a solid body containing a dilute "suspension" of voids, which is a more appropriate description for the structure of a meteorite samples than the regolith model. Eq. \ref{eq7}, together with the measured heat conductivities of the meteorites from \citet{Opeil2010}, yield the bulk heat conductivities of the non-porous materials used in this work,
\begin{align}
&\lambda_{solid,S}\,  =  \,\;   2.18 \,                                                    \mathrm{W \, m^{-1} \, K^{-1}}\, \mathrm{,}\nonumber\\[2mm]
&\lambda_{solid,C}(T)                          \,  =  \,   (1.19 \, + \,             2.1 \times10^{-3} \, T[\mathrm{K}] )         \, \mathrm{W \, m^{-1} \, K^{-1}}\, \mathrm{,}\nonumber\\[2mm]
&\lambda_{solid,M}(T)                          \,  =  \,  (12.63 \, + \,            5.1 \times10^{-2} \, T[\mathrm{K}] )        \, \mathrm{W \, m^{-1} \, K^{-1}}\, \mathrm{.}
\label{eq8}
\end{align}
\par
Since the temperature dependence of the specific surface energy of basaltic rocks and meteorites is not known, we used the corresponding value estimated for $\mathrm{SiO_2}$ \citep{Gundlach2012}. Furthermore, an emissivity of $\epsilon \, = \, 1$ was assumed for the S and C class asteroids. The calculations for the M class asteroids were performed with an emissivity of $\epsilon = 0.66$. This value was estimated using Eq. \ref{eq9} together with the measured surface temperature and the albedo of the M class Asteroid Steins (geometric albedo: 0.22; Bond albedo = 0.10; see Sect. \ref{Important properties of the analyzed objects} for details).

\section{Calibration of the heat conductivity model using lunar regolith}\label{Calibration of the heat conductivity model using lunar regolith}
\begin{figure}[b!]
\centering
\begin{overpic}[angle=0,width=1\columnwidth]{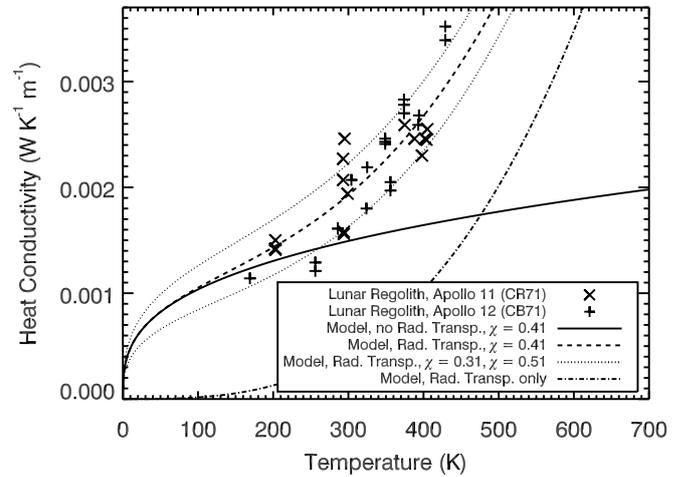}
\end{overpic}
\caption{Calibration of the heat conductivity model with lunar regolith. The measured heat conductivity of lunar regolith is plotted as crosses \citep[Apollo-11;][]{Cremers1971a} and pluses \citep[Apollo-12;][]{Cremers1971b}. Our heat conductivity model of granular material (Eqs. \ref{eq1} - \ref{eq6}) was fit to the data in order to estimate the model parameter $\chi$, which takes the irregular shape of the particles into account. The solid and the dashed curves show the model results for $\chi \, = \, 0.41$ without and with radiative heat transport, respectively. For comparison, the model results for $\chi \, = \, 0.31$ and $\chi \, = \, 0.51$ are also shown as dotted curves. For these calculations, a volume filling factor of $\phi \, = \, 0.43$ \citep{Cremers1971a, Cremers1971b} and a mean particle radius of $r \, = \, 20\, \mathrm{\mu m}$ \citep{McKay2009} were used. Additionally, the dashed-dotted curve shows the radiative heat conductivity without the network heat conductivity for comparison.}
\label{Fig1}
\end{figure}
In order to take the irregular shape of the regolith particles into account, we determined the parameter $\chi$ in Eq. \ref{eq3} and \ref{eq6} using ground-based measurements of the heat conductivity of lunar regolith and, thus, re-calibrated the heat conductivity model \citep{Gundlach2012} to natural regolith.
\par
For the calibration of the heat conductivity model, we used the measured temperature dependence of the heat conductivity of lunar regolith returned with the Apollo-11 and the Apollo-12 missions \citep{Cremers1971a,Cremers1971b} and compared it to the result of our heat conductivity model (see Sect. \ref{The heat conductivity model}), using the known or estimated material properties of the lunar regolith samples (see Tab. \ref{Table_2a}). For completeness, we should mention that the heat conductivity of lunar regolith was also estimated from radio emission measurements of the lunar surface \citep{Krotikov1963}. However, the heat conductivity derived from radio emission measurements is approximately one magnitude higher compared to the values measured in the laboratory. Because of the better accuracy of the laboratory experiments \citep[the error of the radio emission measurements was estimated to be $50\,\%$ of the measured value,][]{Krotikov1963}, we decided to use the results obtained by \citet{Cremers1971a} and the results obtained by \citet{Cremers1971b}.
\par
Fig. \ref{Fig1} shows the temperature dependence of the measured heat conductivity of lunar regolith returned with the Apollo-11 \citep[crosses;][]{Cremers1971a} and Apollo-12 missions \citep[pluses;][]{Cremers1971b} together with the result of the heat conductivity model described in Eqs. \ref{eq1}-\ref{eq6} without (solid curve) and with radiative heat transport (dashed and dotted curves). The model parameters used here are summarized in Tab. \ref{Table_2a}.
\par
The volume filling factor $\phi \, = \, 0.43$ was estimated using the measured density of the lunar regolith, $\rho_{\rm regolith} \, = \, 1300 \, \mathrm{kg \, m^{-3}}$ \citep{Cremers1971a,Cremers1971b}, and the assumption that the density of the solid regolith material is represented by the density of the lunar meteorites, $\rho_{\rm solid} \, = \, 3040 \, \mathrm{kg \, m^{-3}}$ \citep{Warren2001}. For the heat conductivity of the solid material, we used the two limiting values $\lambda_{\rm solid} = 2.18\, \mathrm{W \, m^{-1} \, K^{-1}}$ \citep{Opeil2010} and $\lambda_{\rm solid} = 1.37 \,\mathrm{W \, m^{-1} \, K^{-1}}$ for $\rm SiO_2$ \citep{Ratcliffe1963}. We obtain for the only unknown parameter in the model, $\chi$, which describes the reduction of the heat conductivity when spherical grains are replaced by irregular grains, values of $\chi =  0.31 \pm 0.01$ and $\chi = 0.50 \pm 0.02$, respectively. Thus, we estimate a mean thermal conductivity of the regolith material of $\lambda_{\rm solid} = 1.78\, \mathrm{W \, m^{-1} \, K^{-1}}$ and a mean reduction factor of $\chi = 0.41 \pm 0.02 \,({\rm stat}) \pm 0.10 \, ({\rm syst})$. Fig. \ref{Fig1} also shows that most of the heat conductivity measurements fall within a band of  $\Delta\chi = \pm 0.10$.
\par
One can recognize that the radiative heat-conductivity term is not negligible for temperatures above $\sim200 \, \mathrm{K}$. However, the first term on the rhs of Eq. \ref{eq1}, which describes the heat conductivity through the particle network, still contributes considerably to the heat conductivity even for the highest temperatures in Fig. \ref{Fig1} so that a calibration of the heat conductivity model is possible.
\par
We also allowed a variation of the scaling parameter $e_1$ in Eq. \ref{eq2} and tested whether the agreement between the thermal conductivity measurements of lunar regolith and the model can be further improved. We get a slightly better reduced chi-square if we use $\chi =  0.37 \pm 0.02$ and $e_1 = 1.66 \pm 0.21$. As this agrees with the model with only one free parameter within $\sim 1.5$ standard deviations, we henceforth used $\chi =  0.41 \pm 0.02$ and $e_1 = 1.34 \pm 0.01$.

\section{Important properties of the analyzed objects}\label{Important properties of the analyzed objects}
Tab. \ref{table_1} summarises important properties of the analyzed objects, including the spectral classes of the objects, the diameters $D$ of the objects, the gravitational acceleration $g$ at the surface of the objects, the measured thermal inertiae $\Gamma$ of their surface materials, their surface temperatures $T$ during observations, and the estimated mean grain radii $r$ of the surface regolith. The asteroids Lutetia and Cybele can be assigned either to the spectral class C or M. In this work, these asteroids are treated as C-class objects. The diameters of the analyzed objects were either taken from literature or from the JPL Small-Body Database\footnote{http://ssd.jpl.nasa.gov/sbdb.cgi}.
\par
The gravitational acceleration at the surface was calculated using the mass and the effective diameter of the object. If no measurement of the mass was available, we calculated the gravitational acceleration assuming spherical objects with a mean density depending on the class of the asteroid: $\rho = 1400 \, \mathrm{kg \, m^{-3}}$ (C), $\rho = 2690 \, \mathrm{kg \, m^{-3}}$ (S) and $\rho = 4700 \, \mathrm{kg \, m^{-3}}$ \citep[M,][]{Britt2002}. For Mercury the gravitational acceleration was taken from the NASA webpage\footnote{http://nssdc.gsfc.nasa.gov/planetary/factsheet/mercuryfact.html}.
\par
For the Moon, the thermal inertia measurements were performed for temperatures between $200\,\mathrm{K}$ and $390\,\mathrm{K}$. Thus, the uncertainty of the grain size estimation is dominated by the temperature variation. For Mercury, the influence of the temperature (ranging from $350\,\mathrm{K}$ to $700\,\mathrm{K}$) on the error of the particle size estimation was also taken into account.
\par
For objects with no direct measurement of the surface temperature, the maximum surface temperature was calculated using the standard thermal model \citep[see][and references therein for details]{Lebofsky1986}. This model assumes spherical, slow rotating objects heated by sunlight and cooled by radiation. Each point on the surface is in instantaneous equilibrium with the Solar radiation and no heat is conducted through the regolith towards the center of the object. This approximation was used due to the fact that all analyzed objects have long rotation periods (rotation period $> 4 \, \mathrm{h}$). Following the standard thermal model, the maximum surface temperature at the subsolar point is given by
\begin{equation}
T \,  =  \, \left[\,\frac{S}{\sigma \, \epsilon \, \eta} \, \left(\, 1 \, - \, A \, \right) \, \frac{a^2_E}{a^2} \,\right]^{1 / 4}
\label{eq9}
\end{equation}
\citep{Lebofsky1989}. Here, $S \,= \, 1367 \, \mathrm{W \, m^{-2}}$ is the Solar constant, $A$ is the Bond albedo of the object, $a_E$ is the semi-major axis of the Earth's orbit, $a$ is the distance of the object to the Sun during observation, $\sigma$ is the Stefan-Boltzmann constant, $\epsilon$ is the emissivity of the surface material and $\eta$ is the so-called beaming parameter of the object, which takes deviations from the standard thermal model (e.g. non isotropic heat radiation) into account. In our work, we use $\eta = 1$ (spherical objects, standard thermal model). The bond albedo $A_{bond}$ was calculated using the phase integral $q$ and the geometric albedo $A_{geom}$, $A_{bond} = q \, A_{geom}$. The values for the geometric albedo and the distance of the object to the Sun at the date of observation were taken from the JPL Small-Body Database$^1$. An approximation for the phase integral for asteroids is given by $q = 0.290 + 0.684 \, G$, with a slope parameter of $G = 0.25$ \citep[see appendix in][]{Bowell1989}. Tab. \ref{Table_3} summarizes the used values and the derived temperatures.

\begin{table*}[p!]
\begin{center}
\vspace{-3.8cm}
\hspace{-5cm}
    \caption{Important properties and estimated grain sizes of the analyzed objects.}\vspace{1mm}
    \scriptsize
    \rotatebox{90}{
    \begin{tabular}{lcr@{$\,$}llr@{$\,$}llr@{$\,$}llrlr@{$\,$}l}
    \hline
    Object & Spectral Class & \multicolumn{3}{c}{\hspace{-1.0cm}D [km]}   & \multicolumn{3}{c}{\hspace{-1.0cm}g [$\mathrm{m \, s^{-2}}$]}  &  \multicolumn{3}{c}{\hspace{-1.0cm}$\Gamma$ [$\mathrm{J \, m^{-2} \, K^{-1} \, s^{-0.5}}$]}  &  \multicolumn{2}{c}{\hspace{-1.5cm}T [K]} &    \multicolumn{2}{l}{\hspace{0.35cm}$r$ [$\mathrm{\mu m}$]}     \\\hline
    Itokawa     &S   &  0.32  & $^{+0.03}_{-0.03}$& \citep{Mueller2005}                & $\big(9.3$  & $^{+0.9}_{-0.9}\big)\times 10^{-5}$ & $^M$  &   750 & $^{+50}_{-300} $        & \citep{Mueller2005}   & 340& \citep{Okada2006} & $\big(2.1$  & $^{+0.3}_{-1.4}\big)\times 10^4$   \\
    1998 $\mathrm{WT_{24}}$ &M& 0.42  & $^{+0.04}_{-0.04}$&  \citep{Busch2008}         & $\big(2.8$  & $^{+0.3}_{-0.3}\big)\times 10^{-4}$ &$^\rho$ &  200 & $^{+100}_{-100}$        & \citep{Harris2007}    & 306&$^h$      & $\big(6.4$ & $^{+11.7}_{-5.1}\big)\times10^2 $    \\
    1999 $\mathrm{JU_{3} }$ &C   &  0.92  & $^{+0.12}_{-0.12}$& \citep{Hasegawa2008}   & $\big(1.8$  & $^{+0.2}_{-0.2}\big)\times 10^{-4}$ & $^\rho$ &   500 & $^{+250}_{-250} $        & \citep{Hasegawa2008}   & 277& $^h$       & $\big(1.8$  & $^{+2.3}_{-1.4}\big)\times 10^4$      \\
    1996 $\mathrm{FG_{3} }$ &C& 1.69  & $^{+0.18}_{-0.12}$&  \citep{Wolters2011}       & $\big(3.3$  & $^{+0.4}_{-0.2}\big)\times 10^{-4}$ & $^\rho$ &   120 & $^{+50}_{-50}  $        & \citep{Wolters2011}   & 281&$^h$   & $\big(9.8 $ & $^{+10.3}_{-7.0}\big)\times10^2   $  \\
    Steins      &M   &   5.3 & $^{+1.4}_{-0.8}  $&  \citep{Keller2010}                 & $\big(3.5$  & $^{+0.9}_{-0.5}\big)\times 10^{-3}$ & $^\rho$ &   125 & $^{+41}_{-41}  $        & \citep{Groussin2011} & 245& \citep{Groussin2011}      & $\big(6.3$   & $^{+6.4}_{-3.8}\big)\times10^2      $     \\
                &   &    & &                   & & & &    &     &  \citep{Leyrat2011}& &    &    &          \\
          &   &    & &                   & & & &    &     &       \citep{Lamy2008} & &        &    &  \\
    Betulia     &C   &   5.4 & $^{+0.6}_{-0.6}  $& \citep{Magri2007}                   & $\big(1.1$  & $^{+0.1}_{-0.1}\big)\times 10^{-3}$ & $^\rho$  &   180 & $^{+60}_{-60}  $        & \citep{Harris2005}    & 307&$^h$     & $\big(1.7$  & $^{+1.4}_{-1.0}\big)\times10^3  $   \\
    Deimos      &C   &  12.4 & $^{+0.4}_{-0.4}  $& \citep{Thomas1989}                  & $\big(3.1$  & $^{+0.1}_{-0.1}\big)\times 10^{-3}$ & $^M$  &    55 & $^{+15}_{-15}  $        & \citep{Lunine1982}    & 148& \citep{Lunine1982}        & $\big(1.3$ & $^{+0.9}_{-0.7}\big)\times10^3      $  \\
    Eros        &S   &  16.84& $^{+0.06}_{-0.06}$& $^b$                                & $\big(6.8$  & $^{+0.2}_{-0.2}\big)\times 10^{-3}$ & $^M$            &   150 & $^{+50}_{-50}  $        & \citep{Mueller2008}   & 252&$^h$        & $\big(2.0$  & $^{+1.6}_{-1.2}\big)\times10^3    $   \\
    Phobos      &C   & 22.2  & $^{+0.3}_{-0.3}  $& \citep{Thomas1989}                  & $\big(5.8$  & $^{+0.1}_{-0.1}\big)\times 10^{-3}$ & $^M$     &    53 & $^{+15}_{-15}  $        & \citep{Lunine1982}    & 148& \citep{Lunine1982} & $\big(1.1$ & $^{+0.9}_{-0.7}\big)\times10^3      $   \\
    Elvira      &S   & 27.2  & $^{+0.9}_{-0.9}  $& $^b$                                & $\big(1.0$  & $^{+0.0}_{-0.0}\big)\times 10^{-2}$ & $^\rho$      &   250 & $^{+150}_{-150}$        & \citep{Delbo2009}     & 197&$^h$    & $\big(1.2$ & $^{+1.9}_{-1.0}\big)\times10^4$          \\
    Bohlinia    &S   & 33.7  & $^{+1.4}_{-1.4}  $& $^b$                                & $\big(1.3$  & $^{+0.1}_{-0.1}\big)\times 10^{-2}$ & $^\rho$     &   135 & $^{+65}_{-65}  $        & \citep{Delbo2009}     & 191&$^h$     & $\big(3.8$ & $^{+4.6}_{-2.9}\big)\times10^3  $   \\
    Unitas      &S   & 46.7  & $^{+2.3}_{-2.3}  $& $^b$                                & $\big(1.8$  & $^{+0.1}_{-0.1}\big)\times 10^{-2}$ & $^\rho$            &   180 & $^{+80}_{-80}  $        & \citep{Delbo2009}     & 203&$^h$       & $\big(4.5$ & $^{+5.0}_{-3.2}\big)\times10^3  $   \\
    Dodona      &M   & 58.4  & $^{+2.8}_{-2.8}  $& $^b$                                & $\big(3.8$  & $^{+0.2}_{-0.2}\big)\times 10^{-2}$ & $^\rho$       &    83 & $^{+68}_{-68}$ $^e     $& \citep{Delbo2009}     & 204&$^h$       & $\big(2.7$ & $^{+10.5}_{-2.5}\big)\times10^2    $    \\
    \hline
    \multicolumn{14}{l}{$D$: diameter of the object.}\\[-0.15cm]
    \multicolumn{14}{l}{$g$: gravitational acceleration at the surface of the object.}\\[-0.15cm]
    \multicolumn{14}{l}{$^a$: the particle size estimation was performed for the spectral class C.}\\[-0.15cm]
    \multicolumn{14}{l}{$^b$: the diameter of the object was taken from the JPL Small-Body Database.}\\[-0.15cm]
    \multicolumn{14}{l}{$^M$: gravitational acceleration calculated using the mass of the object and assuming a spherical body with diameter $D$.}\\[-0.15cm]
    \multicolumn{14}{l}{$^\rho$: gravitational acceleration calculated using different mean densities depending on the class of the asteroid,}\\[-0.22cm]
    \multicolumn{14}{l}{\hspace{0.2cm} $\rho = 1400 \, \mathrm{kg \, m^{-3}}$ (C), $\rho = 2690 \, \mathrm{kg \, m^{-3}}$ (S) and $\rho = 4700 \, \mathrm{kg \, m^{-3}}$ \citep[M,][]{Britt2002} and assuming a spherical body with diameter $D$.}\\[-0.15cm]
    \multicolumn{14}{l}{$^G$: gravitational acceleration taken from http://nssdc.gsfc.nasa.gov/planetary/factsheet/mercuryfact.html.}\\[-0.15cm]
          \multicolumn{14}{l}{$\Gamma$: measured thermal inertia of the surface regolith.}\\[-0.15cm]
        \multicolumn{14}{l}{$T$: temperature of the surface regolith.}\\[-0.15cm]
        \multicolumn{14}{l}{$r$: estimated mean grain radius of the surface regolith.}\\[-0.15cm]
        \multicolumn{14}{l}{$^c$: for the Moon, the thermal inertia measurements were performed for temperatures between $200\,\mathrm{K}$ and $390\,\mathrm{K}$.}\\[-0.22cm]
        \multicolumn{14}{l}{\hspace{0.2cm} This temperature variation dominates the uncertainty of the grain size estimation.}\\[-0.15cm]
        \multicolumn{14}{l}{$^d$: for Mercury, the thermal inertia measurements were performed for temperatures between $350\,\mathrm{K}$ and $700\,\mathrm{K}$.}\\[-0.22cm]
        \multicolumn{14}{l}{\hspace{0.2cm} The influence of the temperature variation on the error of the grain size estimation was also taken into account.}\\[-0.15cm]
        \multicolumn{14}{l}{$^e$: only part of the measured thermal inertia range was used to determine the particle size (see Tab. \ref{table_4}).}\\[-0.15cm]
        \multicolumn{14}{l}{$^f$: no error discussion of the thermal inertia available.}\\[-0.15cm]
        \multicolumn{14}{l}{$^h$: surface temperature calculated using the JPL Small-Body Database data (see Tab. \ref{Table_3}).}\\
    \end{tabular}
    }
     \label{table_1}
\end{center}
\end{table*}

\begin{table*}[p!]
\begin{center}
\vspace{-3.8cm}
\hspace{-5cm}
\begin{center}
    \footnotesize{Table \ref{table_1} (cont.): Important properties and estimated grain sizes of the analyzed objects.}\\[0.15cm]
    \end{center}
    \scriptsize
   \rotatebox{90}{
    \begin{tabular}{lcr@{$\,$}llr@{$\,$}llr@{$\,$}llrlr@{$\,$}l}
    \hline
    Object & Spectral Class & \multicolumn{3}{c}{\hspace{-1.0cm}D [km]}   & \multicolumn{3}{c}{\hspace{-1.0cm}g [$\mathrm{m \, s^{-2}}$]}  &  \multicolumn{3}{c}{\hspace{-1.0cm}$\Gamma$ [$\mathrm{J \, m^{-2} \, K^{-1} \, s^{-0.5}}$]}  &  \multicolumn{2}{c}{\hspace{-1.5cm}T [K]} &    \multicolumn{2}{l}{\hspace{0.35cm}$r$ [$\mathrm{\mu m}$]}     \\\hline
    Nysa        &M   & 70.6  & $^{+4.0}_{-4.0}  $& $^b$                                & $\big(4.6$  & $^{+0.3}_{-0.3}\big)\times 10^{-2}$ & $^\rho$                            &   120 & $^{+40}_{-40}  $        & \citep{Delbo2009}     & 196&$^h$       & $\big(8.3$ & $^{+10.1}_{-4.6}\big)\times10^2  $     \\
    Thyra       &S   & 79.8  & $^{+1.4}_{-1.4}  $& $^b$                                & $\big(3.0$  & $^{+0.1}_{-0.1}\big)\times 10^{-2}$ & $^\rho$             &    63 & $^{+38}_{-38}  $        & \citep{Delbo2009}     & 205&$^h$       & $\big(5.5$ & $^{+9.5}_{-5.0}\big)\times10^2    $     \\
    Pomona      &S   & 80.8  & $^{+1.6}_{-1.6}  $& $^b$                                & $\big(3.0$  & $^{+0.1}_{-0.1}\big)\times 10^{-2}$ & $^\rho$              &    70 & $^{+50}_{-50}  $        & \citep{Delbo2009}     & 191&$^h$       & $\big(9.2$ & $^{+17.8}_{-8.8}\big)\times10^2    $      \\
    Lydia       &M   & 86.1  & $^{+2.0}_{-2.0}  $& $^b$                                & $\big(5.7$  & $^{+0.1}_{-0.1}\big)\times 10^{-2}$ & $^\rho$                             &   135 & $^{+65}_{-65}  $        & \citep{Delbo2009}     & 192&$^h$      & $\big(1.1$ & $^{+1.9}_{-0.9}\big)\times10^3  $       \\
    Ekard       &C   & 90.8  & $^{+4.0}_{-4.0}  $& $^b$                                & $\big(1.8$  & $^{+0.1}_{-0.1}\big)\times 10^{-2}$ & $^\rho$                &   120 & $^{+20}_{-20}  $        & \citep{Delbo2009}     & 238&$^h$        & $\big(1.7$ & $^{+0.6}_{-0.5}\big)\times10^3      $        \\
    Lutetia  &M/C$^a$& 95.8  & $^{+4.1}_{-4.1}  $& $^b$                                & $\big(4.9$  & $^{+0.2}_{-0.2}\big)\times 10^{-2}$ &  $^M$                        &    50 & $^{+25}_{-25}  $        & \citep{Mueller2006}   & 245& \citep{Coradini2011}        & $\big(2.1$ & $^{+3.4}_{-1.7}\big)\times10^2      $     \\
    Herculina   &M   & 222.4 & $^{+4.2}_{-4.2}  $& $^b$                                & $\big(1.2$  & $^{+0.0}_{-0.0}\big)\times 10^{-1}$ & $^M$                           &    15 & $^{+8}_{-8  }$ $^e     $& \citep{Mueller1998}   & 183&$^h$     & $\big(2.8$ & $^{+5.1}_{-1.8}\big)\times10^1        $  \\
    Cybele   &M/C$^a$& 237.3 & $^{+4.2}_{-4.2}  $& $^b$                                & $\big(8.5$  & $^{+0.2}_{-0.2}\big)\times 10^{-2}$ & $^M$               &    15 & $^{+8}_{-8  }$ $^e     $& \citep{Mueller2004}   & 169&$^h$        & $\big(3.1$ & $^{+7.9}_{-2.6}\big)\times10^1        $      \\
    Vesta       &S   & 516   & $^{+24}_{-24}    $& \citep{Thomas1997}                  & $\big(2.6$  & $^{+0.1}_{-0.1}\big)\times 10^{-1}$ &$^M$&    25 & $^{+13}_{-13}$          & \citep{Mueller1998}   & 200&$^h$        & $\big(5.4$ & $^{+13.0}_{-4.7}\big)\times10^1       $ \\
    Pallas      &C   & 545   & $^{+18}_{-18}    $& $^b$                                & $\big(2.1$  & $^{+0.3}_{-0.3}\big)\times 10^{-1}$ &$^M$       &    10 & $^{+5}_{-5}  $ $^e     $& \citep{Mueller1998}   & 176&$^h$        & $\big(9.2$ & $^{+27.8}_{-0.0}\big)\times10^0         $   \\
    Ceres       &C   & 968   & $^{+40}_{-40}    $& \citep{Thomas2005}                  & $\big(2.7$  & $^{+0.1}_{-0.1}\big)\times 10^{-1}$ &$^M$   &    38 & $^{+14}_{-14}$          & \citep{Mueller1998}   & 235& \citep{SaintPe1993}        & $\big(1.2$ & $^{+1.2}_{-0.9}\big)\times10^2         $  \\
    Moon  &S&3475.06& $^{+0.06}_{-0.06}   $& \citep{Bills1977}                         & $\big(1.6$  & $^{+0.0}_{-0.0}\big)\times 10^{0}$ & $^M$  &    43 &              $^f     $& \citep{Wesselink1948} & 200 - 390& \citep{Shorthill1972}        & $\big(4.8$ & $^{+11.5}_{-2.7}\big)\times10^1            $   \\
    Mercury &S & 4880    & $^{+1}_{-1}        $& \citep{Yoder1995}                     & $\big(3.7$  & $^{+0.0}_{-0.0}\big)\times 10^{0}$ & $^G$ &  80 & $^{+50}_{-40}  $    & \citep{Yan2006} & 350 - 700& \citep{Yan2006}    & $\big(2.2$ & $^{+4.6+9.5}_{-1.8}\big)\times10^1         $     \\
    &   &    & &                   & & & &    &     &  \citep{Chase1976} & &  \citep{Vas1999}        &  & \\
    \hline
    \multicolumn{14}{l}{$D$: diameter of the object.}\\[-0.15cm]
    \multicolumn{14}{l}{$g$: gravitational acceleration at the surface of the object.}\\[-0.15cm]
    \multicolumn{14}{l}{$^a$: the particle size estimation was performed for the spectral class C.}\\[-0.15cm]
    \multicolumn{14}{l}{$^b$: the diameter of the object was taken from the JPL Small-Body Database.}\\[-0.15cm]
    \multicolumn{14}{l}{$^M$: gravitational acceleration calculated using the mass of the object and assuming a spherical body with diameter $D$.}\\[-0.15cm]
    \multicolumn{14}{l}{$^\rho$: gravitational acceleration calculated using different mean densities depending on the class of the asteroid,}\\[-0.22cm]
    \multicolumn{14}{l}{\hspace{0.2cm} $\rho = 1400 \, \mathrm{kg \, m^{-3}}$ (C), $\rho = 2690 \, \mathrm{kg \, m^{-3}}$ (S) and $\rho = 4700 \, \mathrm{kg \, m^{-3}}$ \citep[M,][]{Britt2002} and assuming a spherical body with diameter $D$.}\\[-0.15cm]
    \multicolumn{14}{l}{$^G$: gravitational acceleration taken from http://nssdc.gsfc.nasa.gov/planetary/factsheet/mercuryfact.html.}\\[-0.15cm]
          \multicolumn{14}{l}{$\Gamma$: measured thermal inertia of the surface regolith.}\\[-0.15cm]
        \multicolumn{14}{l}{$T$: temperature of the surface regolith.}\\[-0.15cm]
        \multicolumn{14}{l}{$r$: estimated mean grain radius of the surface regolith.}\\[-0.15cm]
        \multicolumn{14}{l}{$^c$: for the Moon, the thermal inertia measurements were performed for temperatures between $200\,\mathrm{K}$ and $390\,\mathrm{K}$.}\\[-0.22cm]
        \multicolumn{14}{l}{\hspace{0.2cm} This temperature variation dominates the uncertainty of the grain size estimation.}\\[-0.15cm]
        \multicolumn{14}{l}{$^d$: for Mercury, the thermal inertia measurements were performed for temperatures between $350\,\mathrm{K}$ and $700\,\mathrm{K}$.}\\[-0.22cm]
        \multicolumn{14}{l}{\hspace{0.2cm} The influence of the temperature variation on the error of the grain size estimation was also taken into account.}\\[-0.15cm]
        \multicolumn{14}{l}{$^e$: only part of the measured thermal inertia range was used to determine the particle size (see Tab. \ref{table_4}).}\\[-0.15cm]
        \multicolumn{14}{l}{$^f$: no error discussion of the thermal inertia available.}\\[-0.15cm]
        \multicolumn{14}{l}{$^h$: surface temperature calculated using the JPL Small-Body Database data (see Tab. \ref{Table_3}).}\\
    \end{tabular}
    }
\end{center}
\end{table*}

\begin{table*}[p!]
\begin{center}
    \scriptsize
    \caption{Parameters used for computation of the surface temperature.}\vspace{1mm}
    \begin{tabular}{lccccc}
        \bottomrule
        Object & $A_{geom}$ &  $A_{Bond}$ & Date of Observation & $a$ [$\mathrm{AU}$] & $T$ [$\mathrm{K}$] \\
        \midrule
        1998 $\mathrm{WT_{24}}$& 0.56                       & 0.26 & 2001-12-04 & 1.01 &  339$^\dag$         \\
        1999 $\mathrm{JU_{3} }$& \hspace{1.3mm}0.06$^\ast$  & 0.03 & 2007-05-16 & 1.41 &  277         \\
        1996 $\mathrm{FG_{3} }$& \hspace{1.6mm}0.04$^\star$ & 0.02 & 2011-01-19 & 1.38 &  281         \\
        Betulia                & 0.08                       & 0.04 & 2002-06-02 & 1.14 &  307         \\
        Eros                   & 0.25                       & 0.12 & 1998-06-29 & 1.62 &  252         \\
        Elvira                 & 0.28                       & 0.13 & 1983-07-28 & 2.64 &  197         \\
        Bohlinia               & 0.20                       & 0.09 & 1983-08-09 & 2.88 &  191         \\
        Unitas                 & 0.21                       & 0.10 & 1983-07-31 & 2.17 &  219         \\
        Dodona                 & 0.16                       & 0.07 & 1983-07-11 & 2.55 &  226$^\dag$         \\
        Nysa                   & 0.55                       & 0.25 & 1983-07-27 & 2.47 &  218$^\dag$         \\
        Thyra                  & 0.27                       & 0.12 & 1983-04-28 & 2.45 &  205         \\
        Pomona                 & 0.26                       & 0.12 & 1983-07-31 & 2.81 &  191         \\
        Lydia                  & 0.18                       & 0.08 & 1983-06-25 & 2.87 &  213$^\dag$         \\
        Ekard                  & 0.05                       & 0.02 & 1983-06-13 & 1.92 &  238         \\
        Herculina              & 0.17                       & 0.08 & 1975-01-01 & 3.13 &  183         \\
        Cybele                 & 0.05                       & 0.02 & 1985-12-10 & 3.80 &  169         \\
        Vesta                  & 0.42                       & 0.19 & 1970-01-01 & 2.46 &  200         \\
        Pallas                 & 0.16                       & 0.07 & 1970-01-01 & 3.42 &  176         \\
        \bottomrule
        \multicolumn{6}{l}{$A_{geom}$: geometric albedo.}\\[-0.15cm]
        \multicolumn{6}{l}{$A_{bond}$: Bond albedo.}\\[-0.15cm]
        \multicolumn{6}{l}{$a$: distance to the Sun during observations.}\\[-0.15cm]
        \multicolumn{6}{l}{$T$: surface temperature at the subsolar point.}\\[-0.15cm]
        \multicolumn{6}{l}{$^\dag$: temperature calculated using an emissivity of $\epsilon = 0.66$ (M class asteroids).}\\[-0.15cm]
        \multicolumn{6}{l}{$^\ast$: geometric albedo from \citet{Hasegawa2008}.}\\[-0.15cm]
        \multicolumn{6}{l}{$^\star$: geometric albedo from \citet{Leon2011}.}\\
    \end{tabular}
     \label{Table_3}
     \end{center}
\end{table*}

\section{Determination of the regolith particle size}\label{Determination of the regolith particle size}
\begin{figure}[b!]
\centering
\begin{overpic}[angle=0,width=1.0\columnwidth]{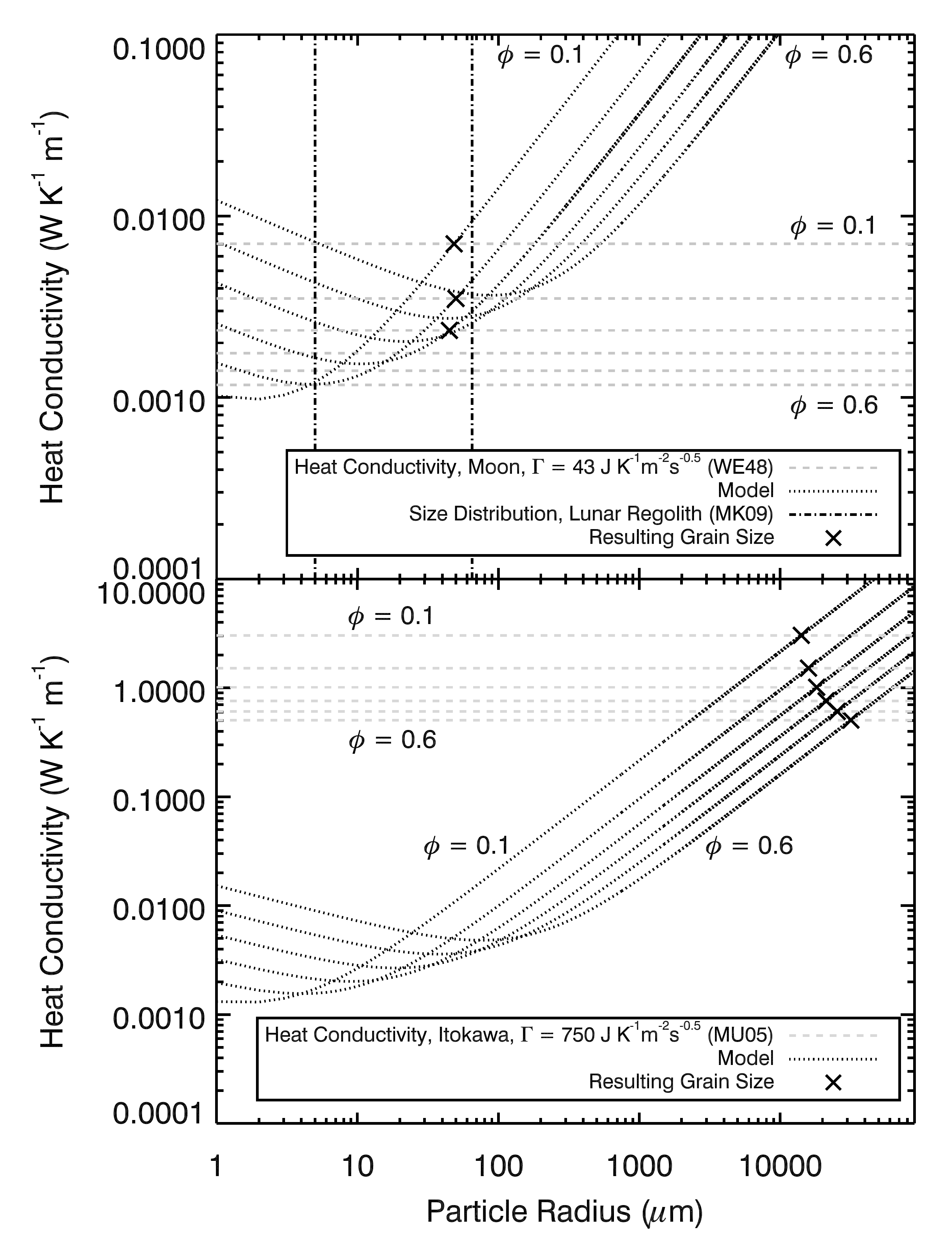}
\put(19,92.5){\Large a)}
\put(19,48.8){\Large b)}
\end{overpic}
\caption{Grain size estimation for the surface regolith of the Moon and asteroid (25143) Itokawa. Results of the regolith heat conductivity model (dotted curves) are shown for the lunar surface (a) and the surface of asteroid (25143) Itokawa (b) for different packing fractions of the regolith particles, $\phi \, = \, 0.1$ to $\phi \, = \, 0.6$, in intervals of $\Delta \phi \, = \, 0.1$. The heat conductivities derived from the thermal inertia measurements, $\Gamma \, = \, 43 \, \mathrm{J \, K^{-1}\, m^{-2}\,s^{-0.5}}$ \citep[Moon;][]{Wesselink1948} and $\Gamma \, = \, 750^{+50}_{-300} \, \mathrm{J \, K^{-1}\, m^{-2}\,s^{-0.5}}$ \citep[Itokawa;][]{Mueller2005}, are shown by the dashed lines, also for $\phi \, = \, 0.1$ to $\phi \, = \, 0.6$. The resulting grain sizes are $r \,=\,  48^{+115}_{-27}\, \mathrm{\mu m}$ for the Moon and $r \,=\, 21^{+3}_{-14}\, \mathrm{mm}$ for asteroid (25143) Itokawa (crosses). Here, the size errors for the lunar regolith derive from the temperature variations of the lunar surface during the measurements, whereas the errors for (25143) Itokawa stem from the uncertainties of the measured thermal inertia. For comparison, the 10\% and 90\% values of the size distribution of the lunar regolith samples returned with the Apollo-11 mission \citep{McKay2009} are also shown (dashed-dotted vertical lines). The model parameters used for the derivation of the regolith grain sizes are summarized in Tabs. \ref{Table_2} and \ref{table_1}.}
\label{Fig2}
\end{figure}

\begin{figure}[b!]
\centering
\begin{overpic}[angle=180,width=1.0\columnwidth]{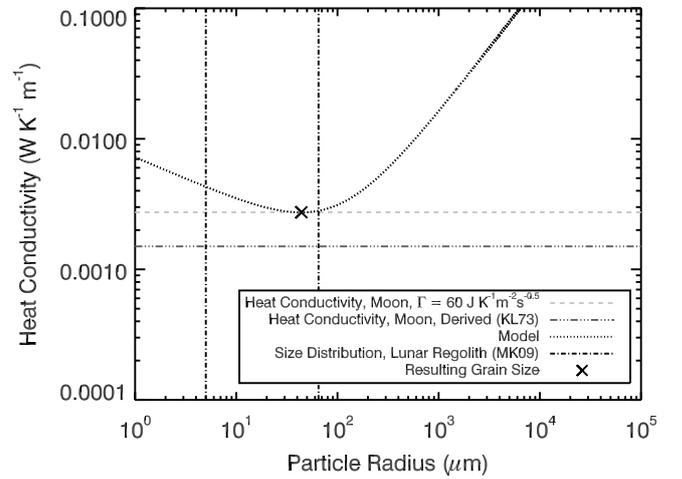}
\end{overpic}
\caption{Grain size estimation performed for the surface regolith of the Moon using a constant value for the volume filling factor of $\phi \, = \, 0.5$. We varied the thermal inertia in order to find a match between the corresponding heat conductivity and the heat conductivity resulting from the model. For a thermal inertia of $\Gamma\, = \, 60 \, \mathrm{J \, m^{-2} \, K^{-1} \, s^{-0.5}}$, a match between the measured heat conductivity, $\lambda \, = \, 2.7 \times 10^{-3}\, \mathrm{W \, m^{-1} \, K^{-1}}$ (dashed line) and the modeled heat conductivity (dotted curve) was found. The comparison of the heat conductivities yields a mean grain radius of $r\, = \, 44 \, \mathrm{\mu m}$. For comparison, the 10\% and 90\% values of the size distribution of the lunar regolith samples returned with the Apollo-11 mission \citep{McKay2009} are also shown (dashed-dotted vertical lines). Additionally, the heat conductivity estimated from the measurement of the infrared radiation of the lunar surface during the Apollo-17 mission is shown,  $\lambda \, = \, 1.5 \times 10^{-3}\, \mathrm{W \, m^{-1} \, K^{-1}}$ (dashed-dotted-dotted line).}
\label{Fig_Moon}
\end{figure}
Applying the thus validated heat-conductivity model to other small Solar System bodies, the mean size of the planetary regolith particles can be estimated. This can be done by comparing the heat conductivities derived from the thermal inertia measurements with those predicted by the model, treating the unknown volume filling factor of the material as a free parameter (see Sect. \ref{Strategy}).
\par
To test this method, the grain size of the lunar regolith was estimated from thermal inertia measurements of the lunar surface (Fig. \ref{Fig2}a). The dashed horizontal lines denote the heat conductivities derived from the measured thermal inertia, $\Gamma \, = \, 43 \, \mathrm{J \, K^{-1}\, m^{-2}\,s^{-0.5}}$ \citep{Wesselink1948}, for filling factors between $\phi \, = \, 0.1$ (uppermost line) and $\phi \, = \, 0.6$ (lowest line). For the Moon, we derived the heat capacity $c$ of the regolith material from the equation for the thermal inertia (see Eq. \ref{eq0}). Using the measured thermal inertia of the lunar surface, $\Gamma = 43 \, \mathrm{J \, m^{-2} \, K^{-1} \, s^{-0.5}}$ \citep{Wesselink1948}, the measured heat conductivity of the lunar samples at 300 K, $\lambda = 2 \times 10^{-3} \,\mathrm{W \, m^{-1} \, K^{-1}}$, and the values for the material and packing density of the lunar regolith shown in Tab. \ref{Table_2a}, we arrive at $c \, = \, 711 \, \mathrm{J \, kg^{-1} \,K^{-1}}$. For the other analyzed objects, the values of the heat capacity summarized in Tab. \ref{Table_2} were used. All relevant values for the estimation of the grain sizes are presented in Tabs. \ref{Table_2} and \ref{table_1}.
\par
The dotted curves in Fig. \ref{Fig2}a represent the model (Eqs. \ref{eq1}-\ref{eq6}) for different volume filling factors, $\phi = 0.1$ to $\phi = 0.6$. The intersection between the model curve and the respective horizontal line yields the mean regolith particle size for each assumed packing density (crosses in Fig. \ref{Fig2}a). As can be seen from Fig. \ref{Fig2}a, one can find such an agreement only for the lower three packing fractions. Taking into account the uncertainties in measuring the thermal inertia of the lunar regolith, we thus arrive at an average grain size of $r \,=\,  48^{+115}_{-27}\, \mathrm{\mu m}$. This falls well into the measured size distribution of the lunar regolith samples returned with the Apollo-11 mission \citep{McKay2009}. In Fig. \ref{Fig2}a we show the 10\% and 90\% values of the size distribution of the lunar regolith by the dashed-dotted vertical lines.
\par
As can be seen, there is no agreement between the model and the derived heat conductivities from thermal inertia measurements for packing fractions above $\phi = 0.3$. This is in contradiction to the measurements of the volume filling factor of the lunar regolith, which indicate $\phi \approx 0.5$ \citep{Mitchell1972}. Thus, we increased the thermal inertia in order to find a match between the derived heat conductivity and the heat conductivity resulting from the model (see Fig. \ref{Fig_Moon}). We only get an agreement if we use a thermal inertia of $\Gamma\, = \, 60 \, \mathrm{J \, m^{-2} \, K^{-1} \, s^{-0.5}}$, which results in a heat conductivity of $\lambda \, = \, 2.7 \times 10^{-3}\, \mathrm{W \, m^{-1} \, K^{-1}}$ (dashed line in Fig. \ref{Fig_Moon}). We thus obtained a regolith grain radius of $r\, = \, 44 \, \mathrm{\mu m}$. The derived grain size is very close to the grain size estimated by varying the volume filling factor of the material, but having volume filling factors lower than $\phi \, = \, 0.5$. However, our value for the heat conductivity is almost a factor of two higher than the value derived from measuring the infrared radiation of the lunar surface during the Apollo-17 mission, $\lambda \, = \, 1.5 \times 10^{-3}\, \mathrm{W \, m^{-1} \, K^{-1}}$ \citep[dashed-dotted-dotted line in Fig. \ref{Fig_Moon};][]{Keihm1973}. \citet{Keihm1973} derived the heat conductivity from a nighttime cool-down curve starting at $\sim 130 \, \mathrm{K}$ and decreasing to $\sim 105 \, \mathrm{K}$. Thus, the discrepancy between our result (derived for $T = 295 \, \mathrm{K}$) and the value estimated by \citet{Keihm1973} is caused by the usage of different temperatures as can be seen in Fig. \ref{Fig1}.
\par
After these calibration and validation steps, we applied our model to various objects for which thermal inertia measurements were available (see Tab. \ref{table_1}) and estimated the grain size of the surface regolith. As the volume filling factor of the regolith is {\it a priori} unknown, we treated its value as a free parameter in this work. Fig. \ref{Fig2}b exemplarily shows our analysis for the surface regolith of the asteroid (25143) Itokawa, target of the sample return mission of the Hayabusa spacecraft. For the measured thermal inertia of $\Gamma \, = \, 750^{+50}_{-300} \, \mathrm{J \, K^{-1}\, m^{-2}\,s^{-0.5}}$ \citep{Mueller2005}, we arrive at a mean grain radius of $r \,=\, 21^{+3}_{-14}\, \mathrm{mm}$, which is in agreement with in-situ observations \citep{Yano2006,Kitazato2008}. The uncertainty of the grain size measurement is completely dominated by the error of the thermal inertia determination. In contrast to the Moon, the heat transport in the regolith on asteroid (25143) Itokawa is governed by radiation so that uncertainties in the material properties are unimportant.
\par
\begin{figure*}[p!]
\centering
\begin{overpic}[angle=180,width=1\columnwidth]{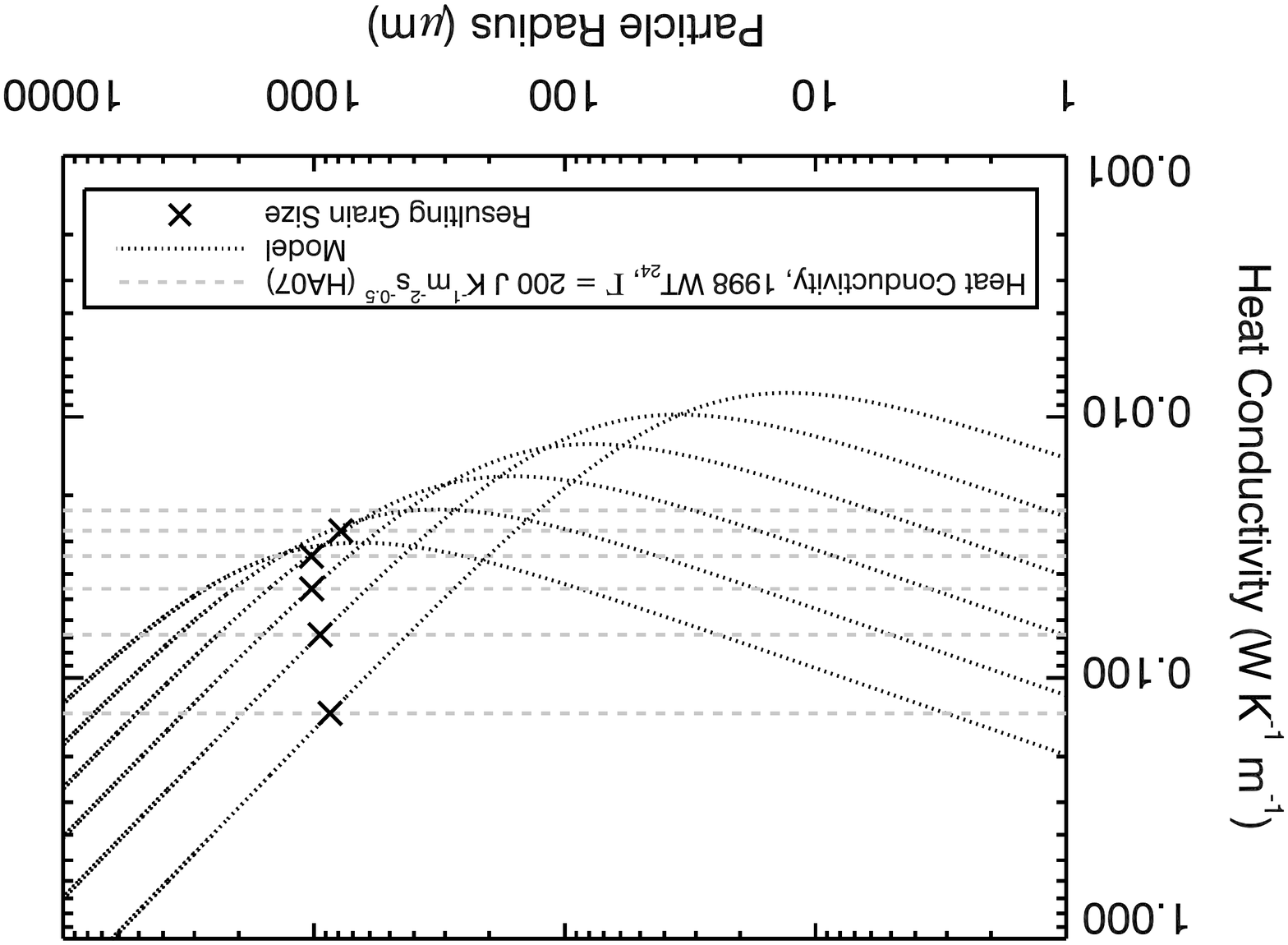}
\put(23,66){\Large a)}
\end{overpic}
\begin{overpic}[angle=180,width=1\columnwidth]{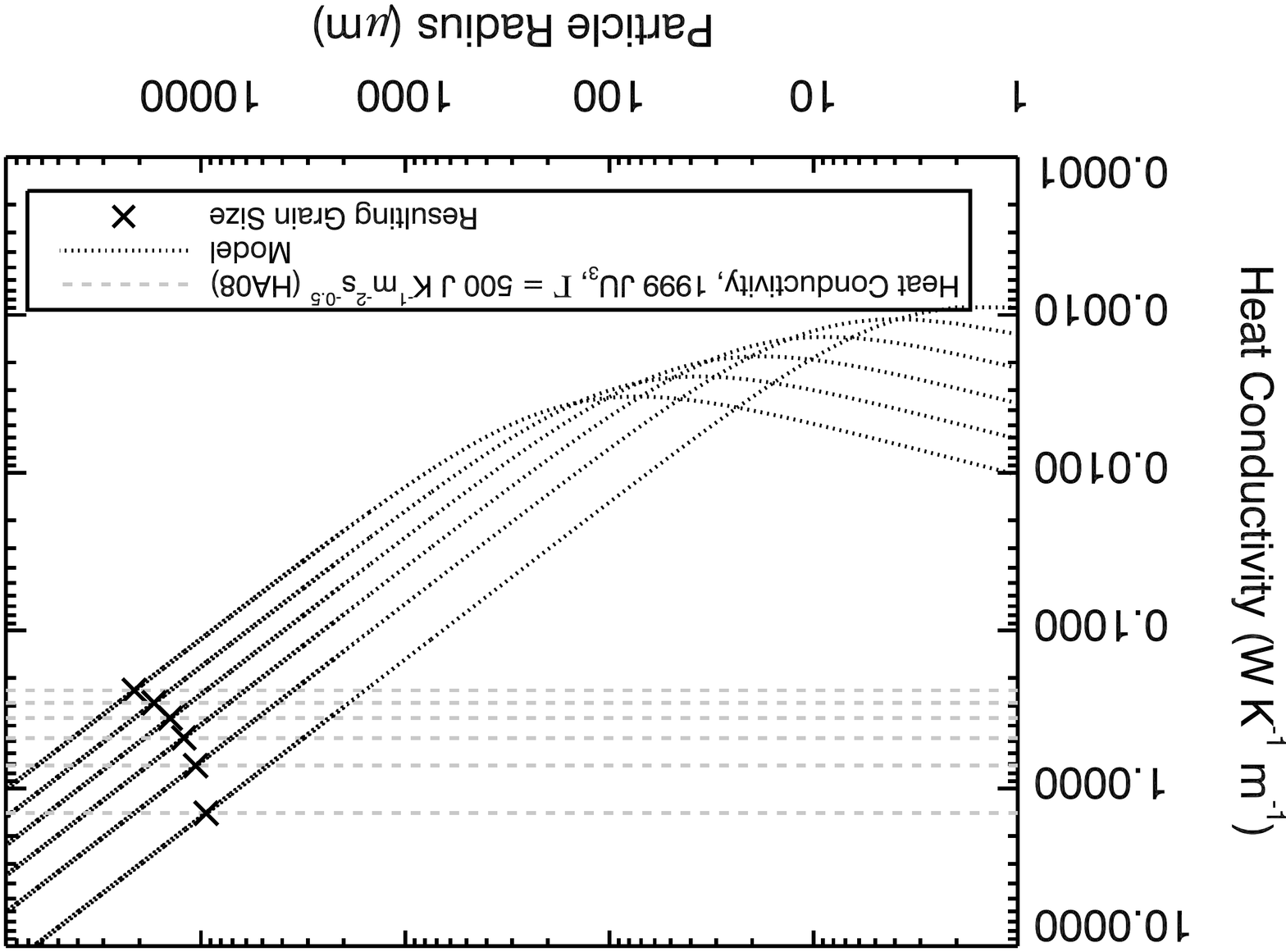}
\put(23,66){\Large b)}
\end{overpic}
\begin{overpic}[angle=180,width=1\columnwidth]{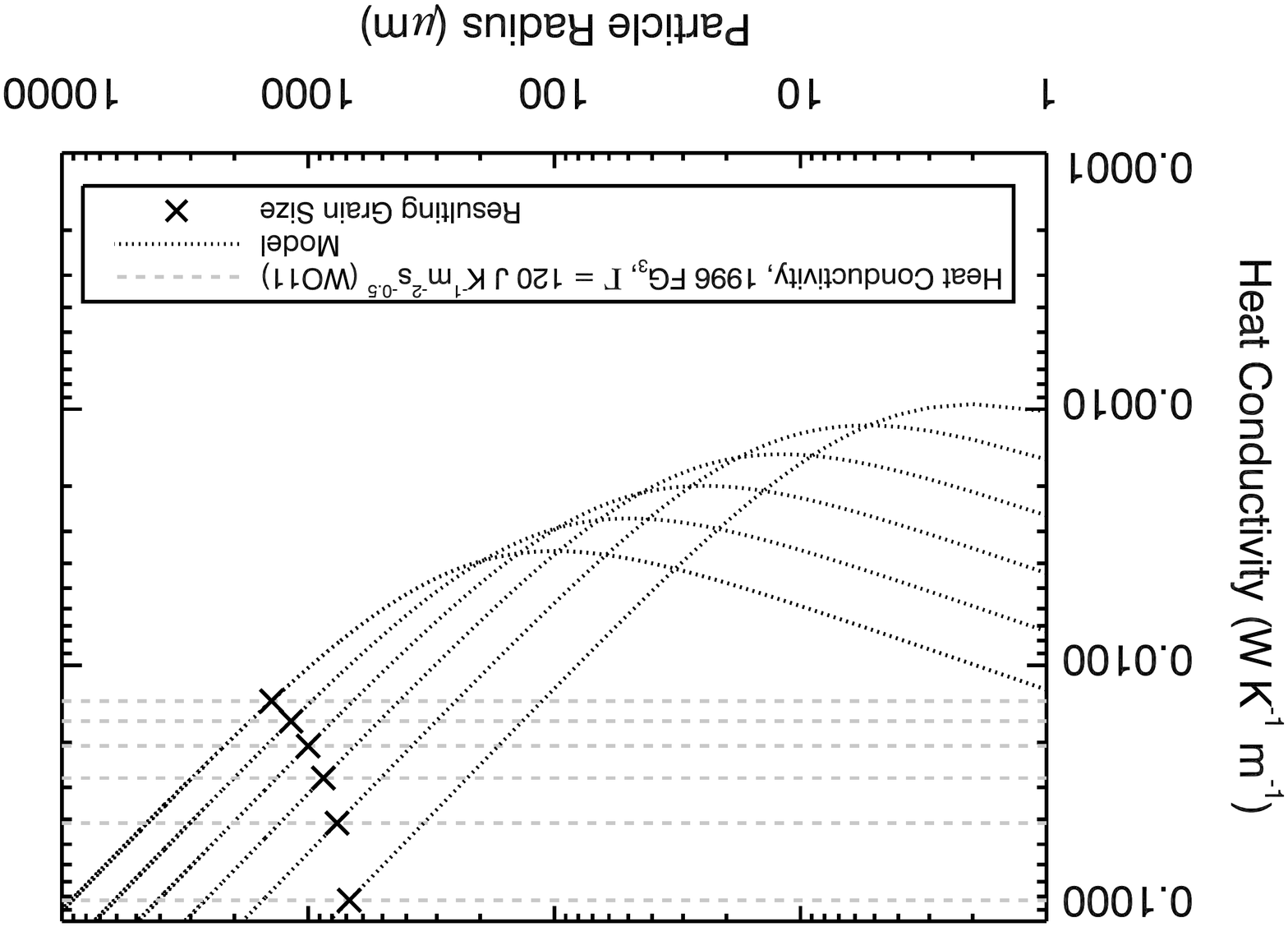}
\put(23,66){\Large c)}
\end{overpic}
\begin{overpic}[angle=180,width=1\columnwidth]{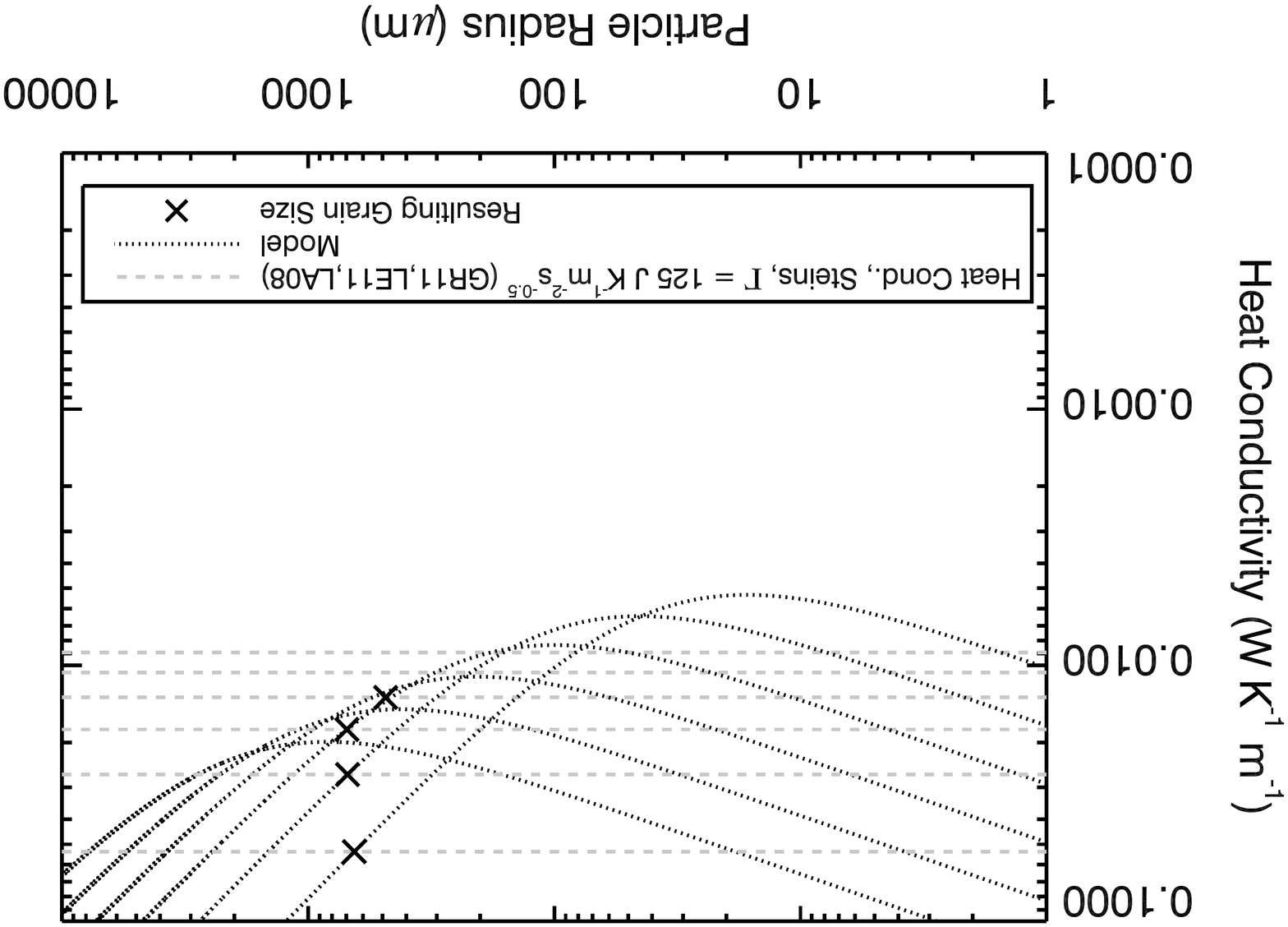}
\put(23,66){\Large d)}
\end{overpic}
\begin{overpic}[angle=180,width=1\columnwidth]{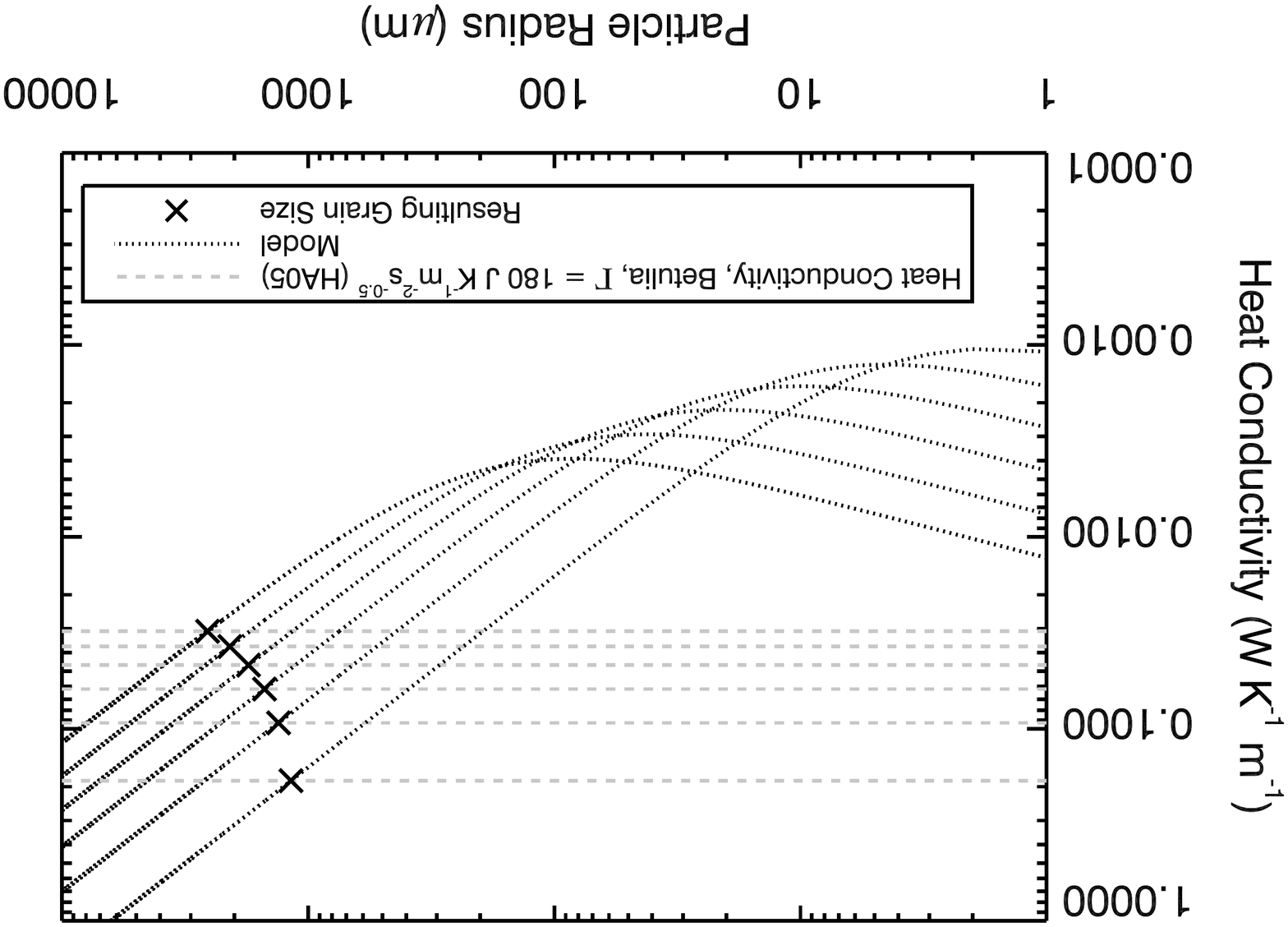}
\put(23,66){\Large e)}
\end{overpic}
\begin{overpic}[angle=180,width=1\columnwidth]{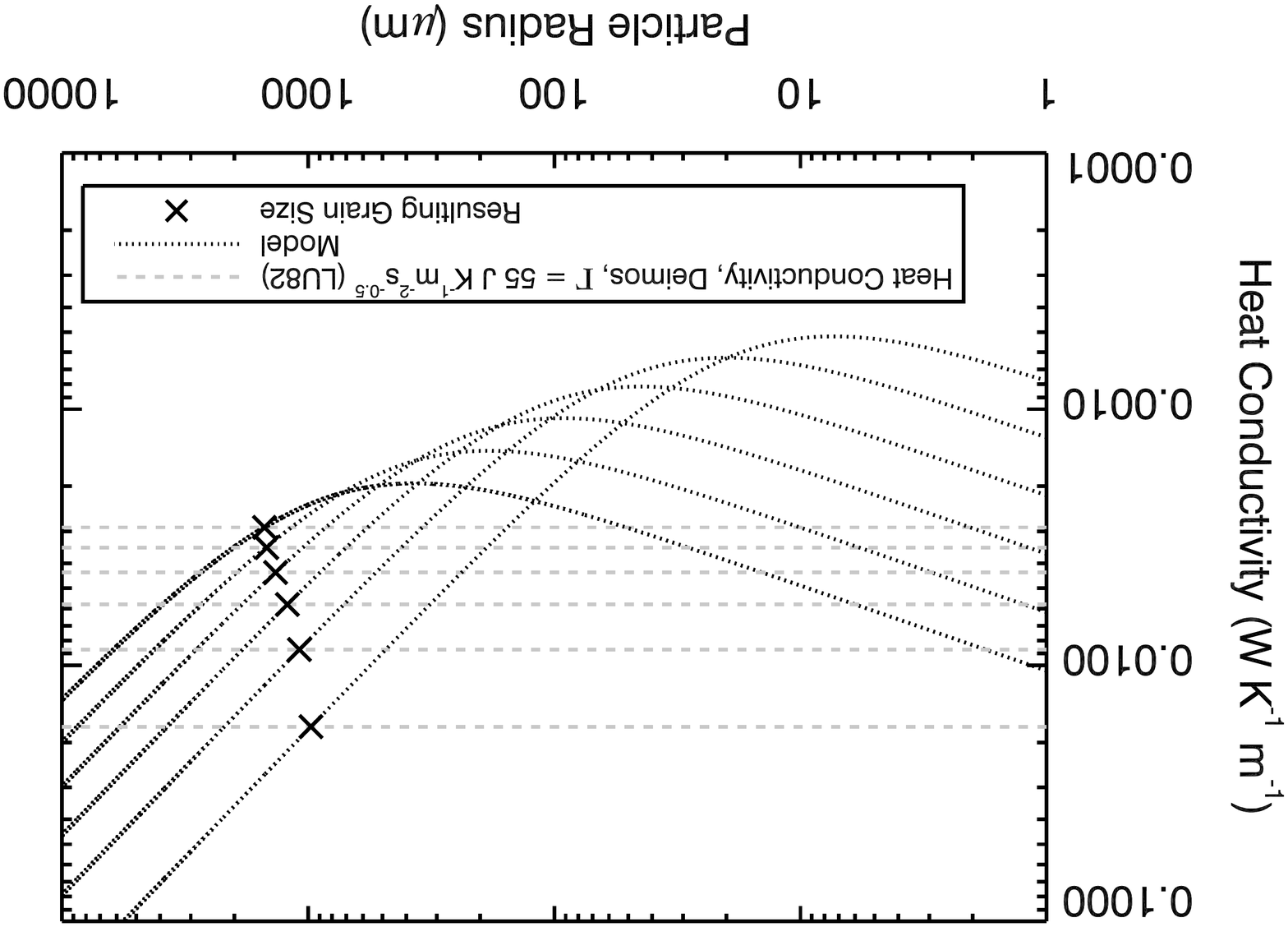}
\put(23,66){\Large f)}
\end{overpic}
\caption{Grain size estimation for the surface regolith of 1998 WT$_{24}$ (a), 1999 JU$_{3}$ (b), 1996 FG$_3$ (c), Steins (d), Betulia (e) and Deimos (f).}
\label{Figcomp1}
\end{figure*}
\begin{figure*}[p!]
\centering
\begin{overpic}[angle=180,width=1\columnwidth]{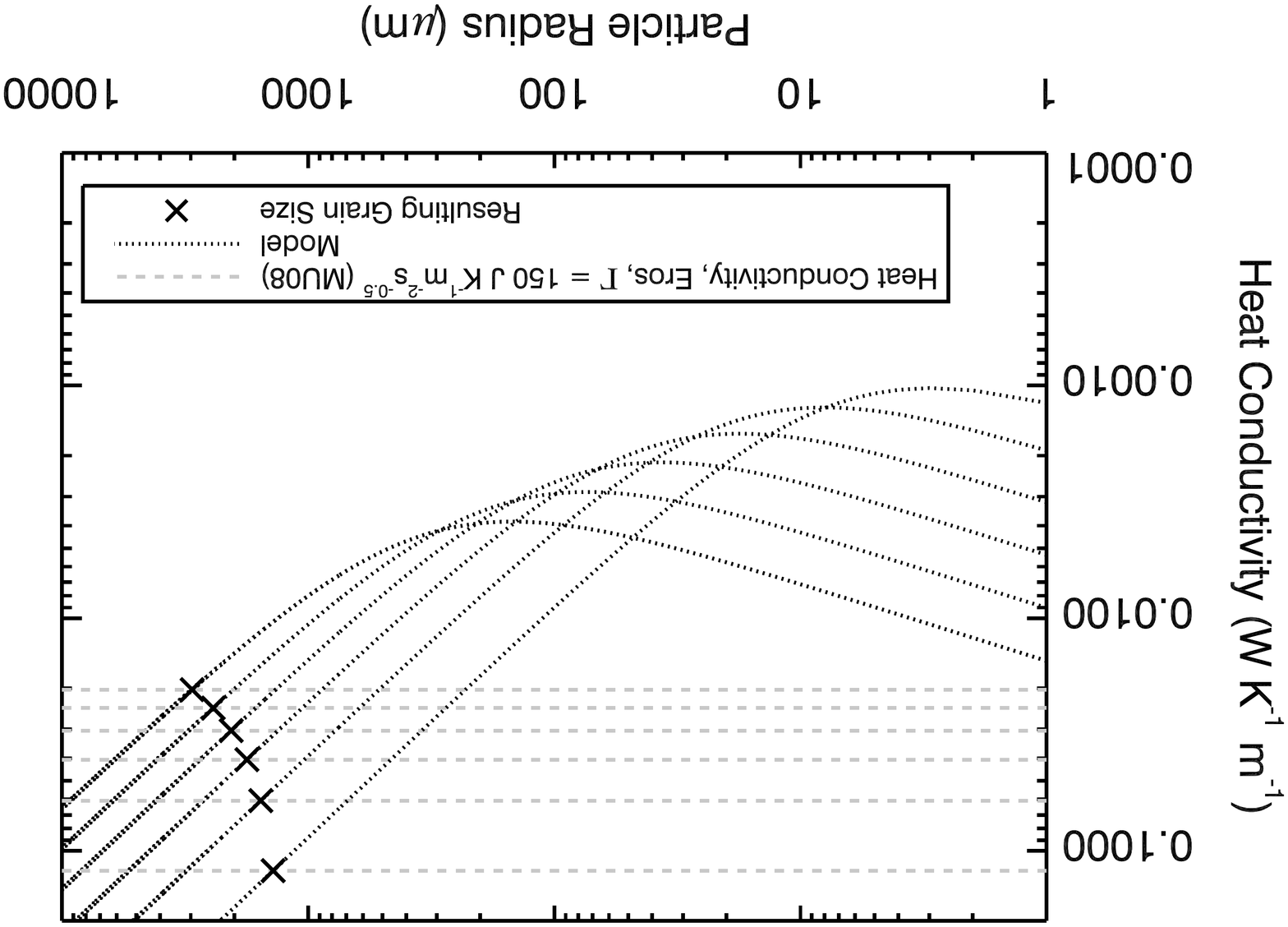}
\put(23,66){\Large a)}
\end{overpic}
\begin{overpic}[angle=180,width=1\columnwidth]{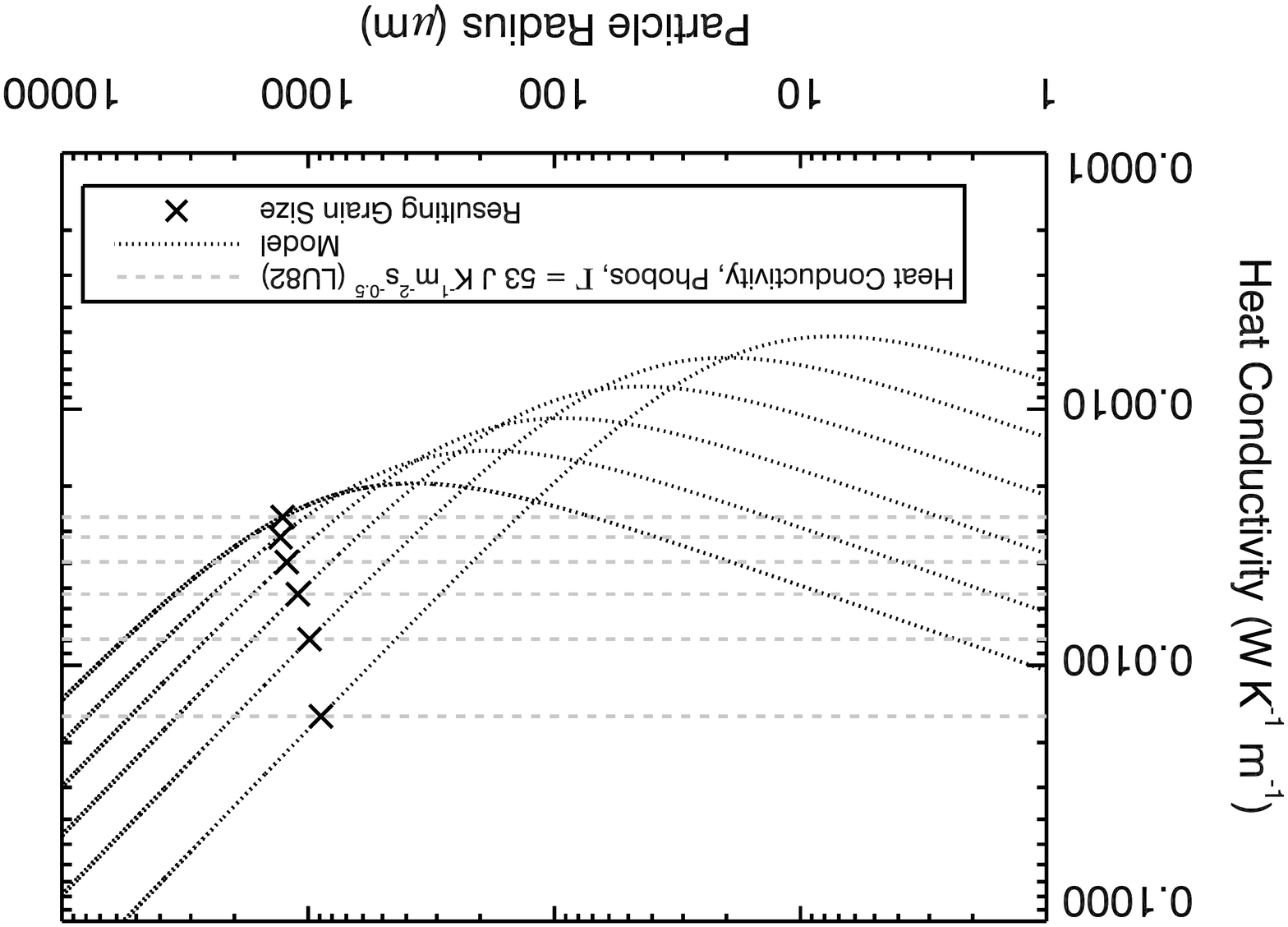}
\put(23,66){\Large b)}
\end{overpic}
\begin{overpic}[angle=180,width=1\columnwidth]{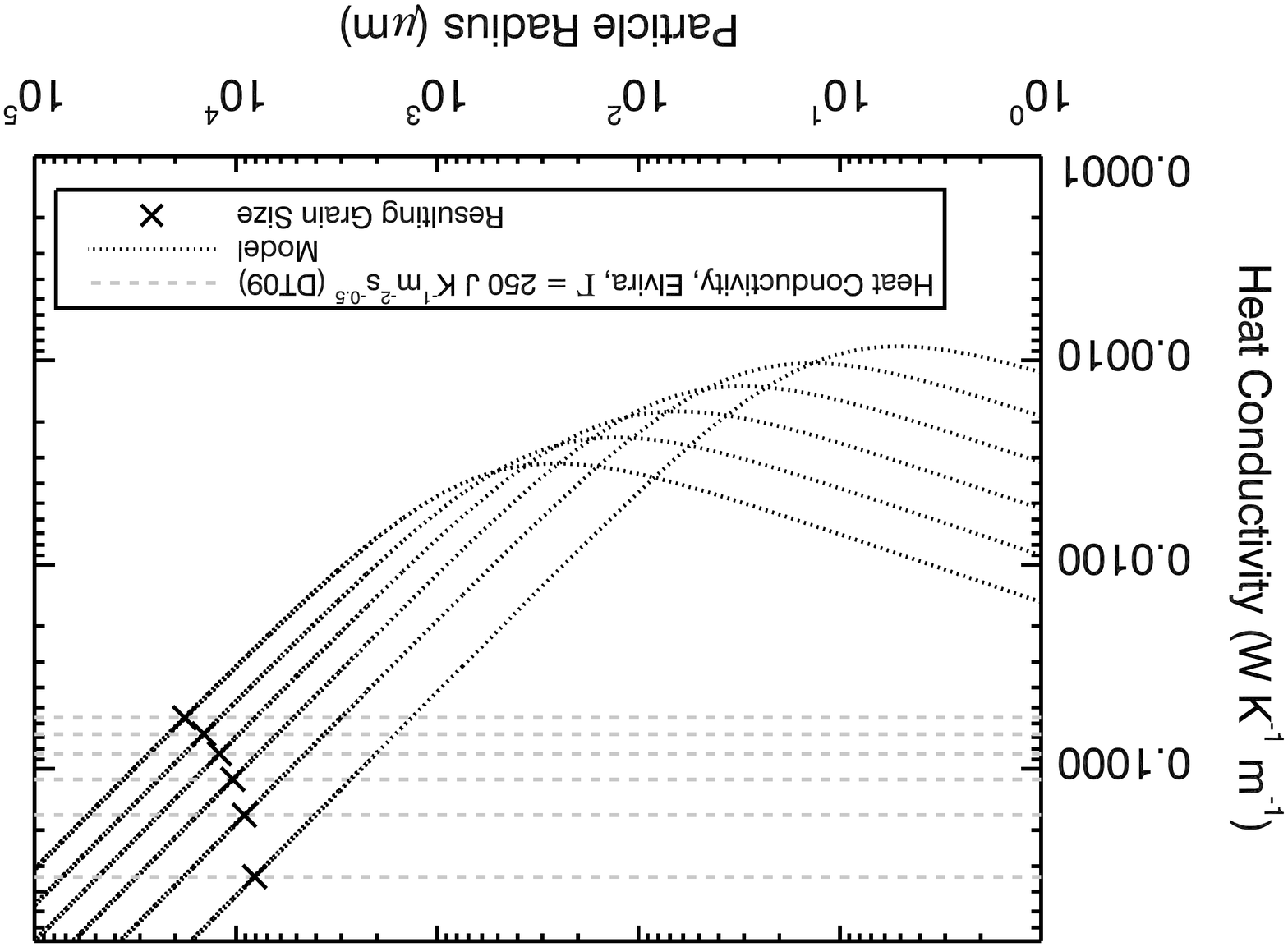}
\put(23,66){\Large c)}
\end{overpic}
\begin{overpic}[angle=180,width=1\columnwidth]{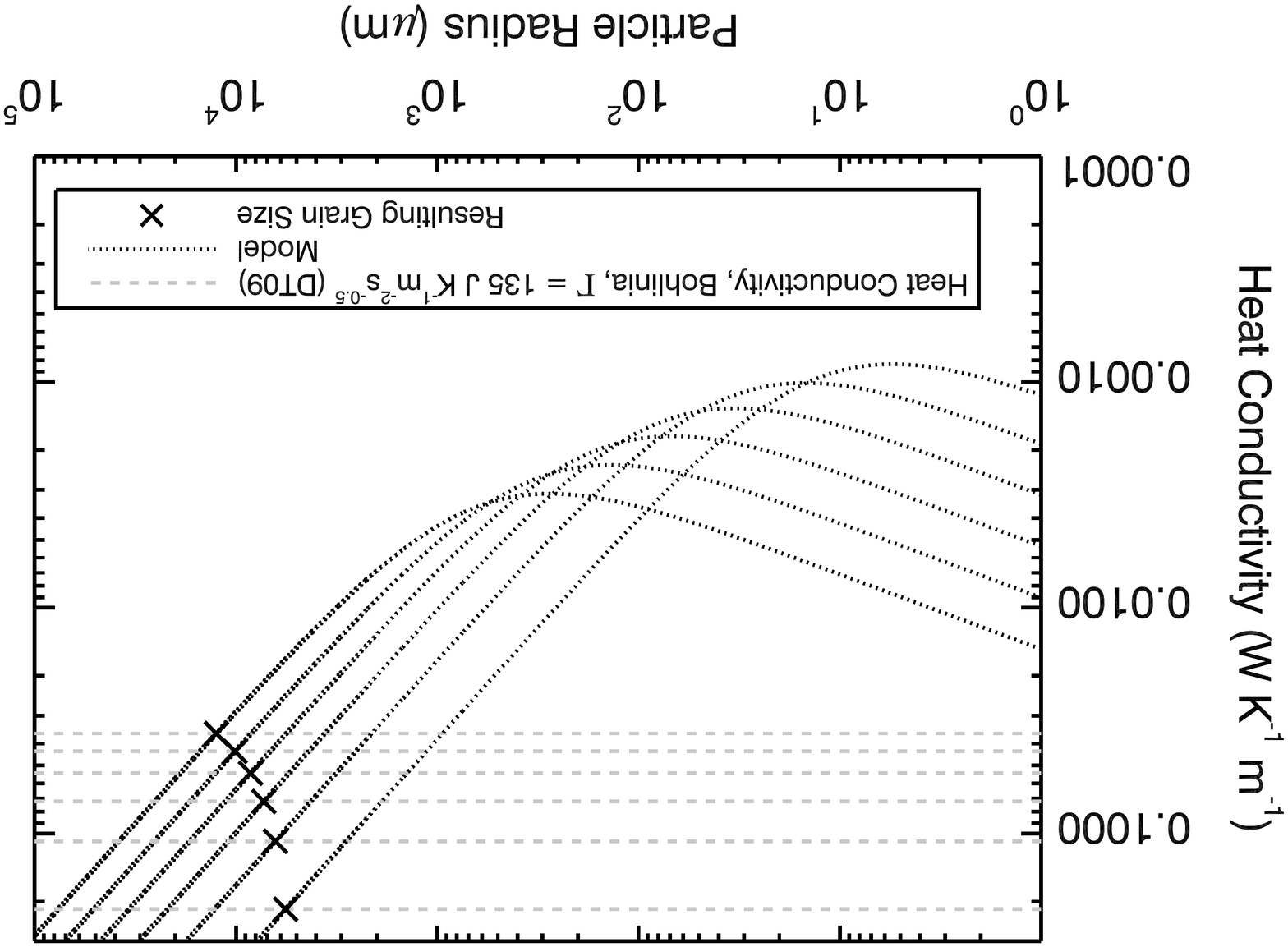}
\put(23,66){\Large d)}
\end{overpic}
\begin{overpic}[angle=180,width=1\columnwidth]{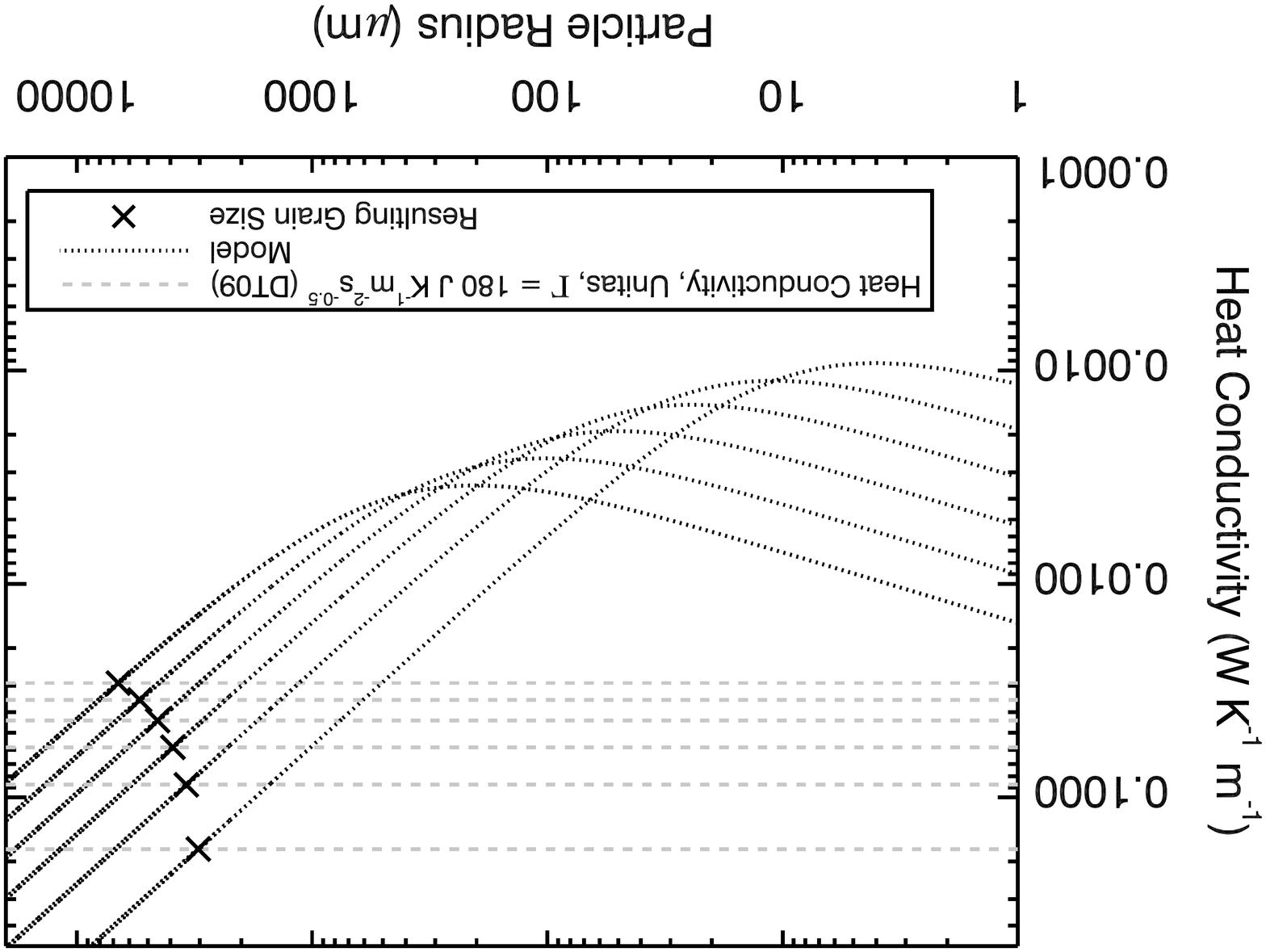}
\put(23,66){\Large e)}
\end{overpic}
\begin{overpic}[angle=180,width=1\columnwidth]{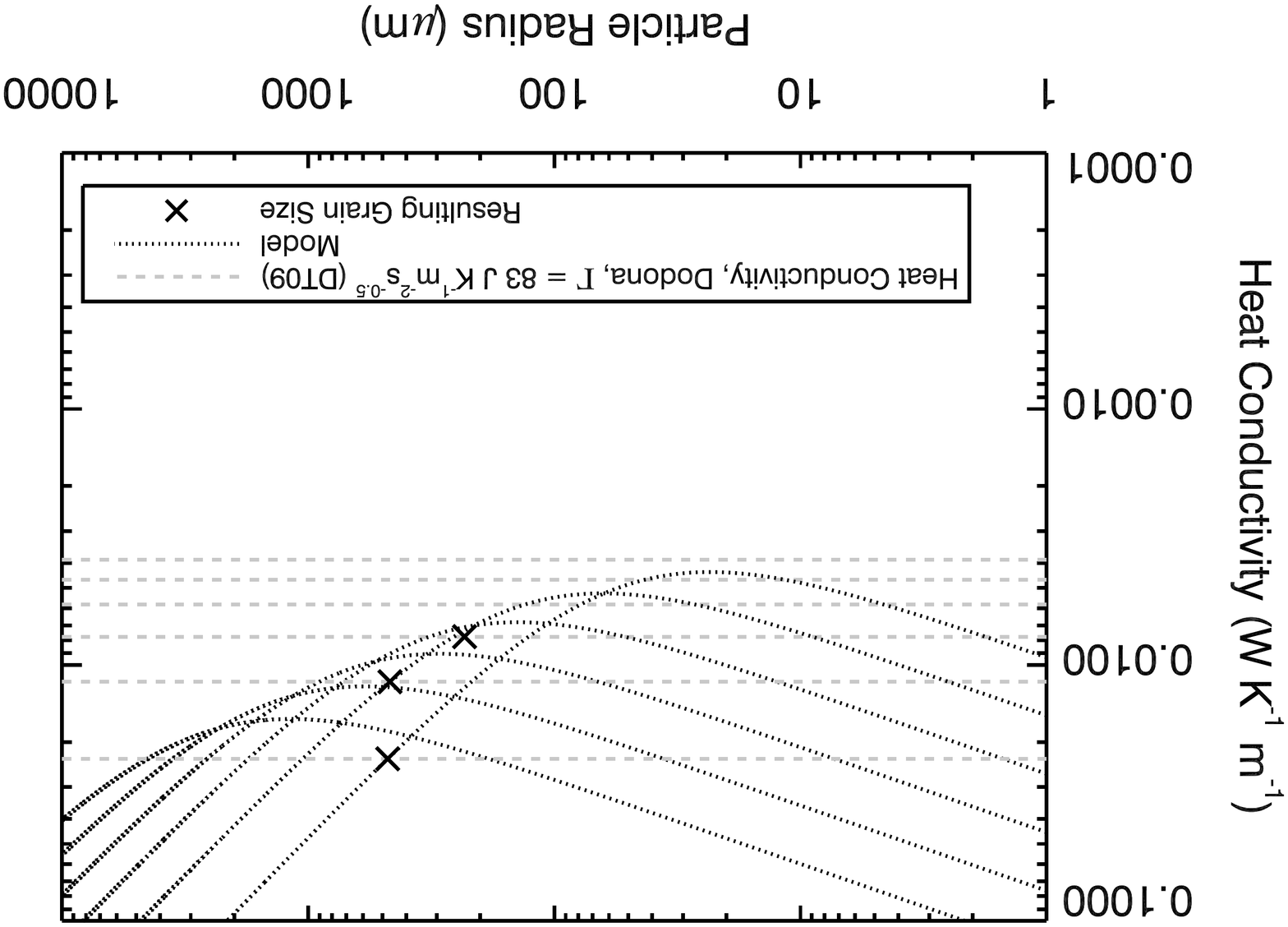}
\put(23,66){\Large f)}
\end{overpic}
\caption{Grain size estimation for the surface regolith of Eros (a), Phobos (b), Elvira (c), Bohlinia (d), Unitas (e) and Dodona (f).}
\label{Figcomp2}
\end{figure*}
\begin{figure*}[p!]
\centering
\begin{overpic}[angle=180,width=1\columnwidth]{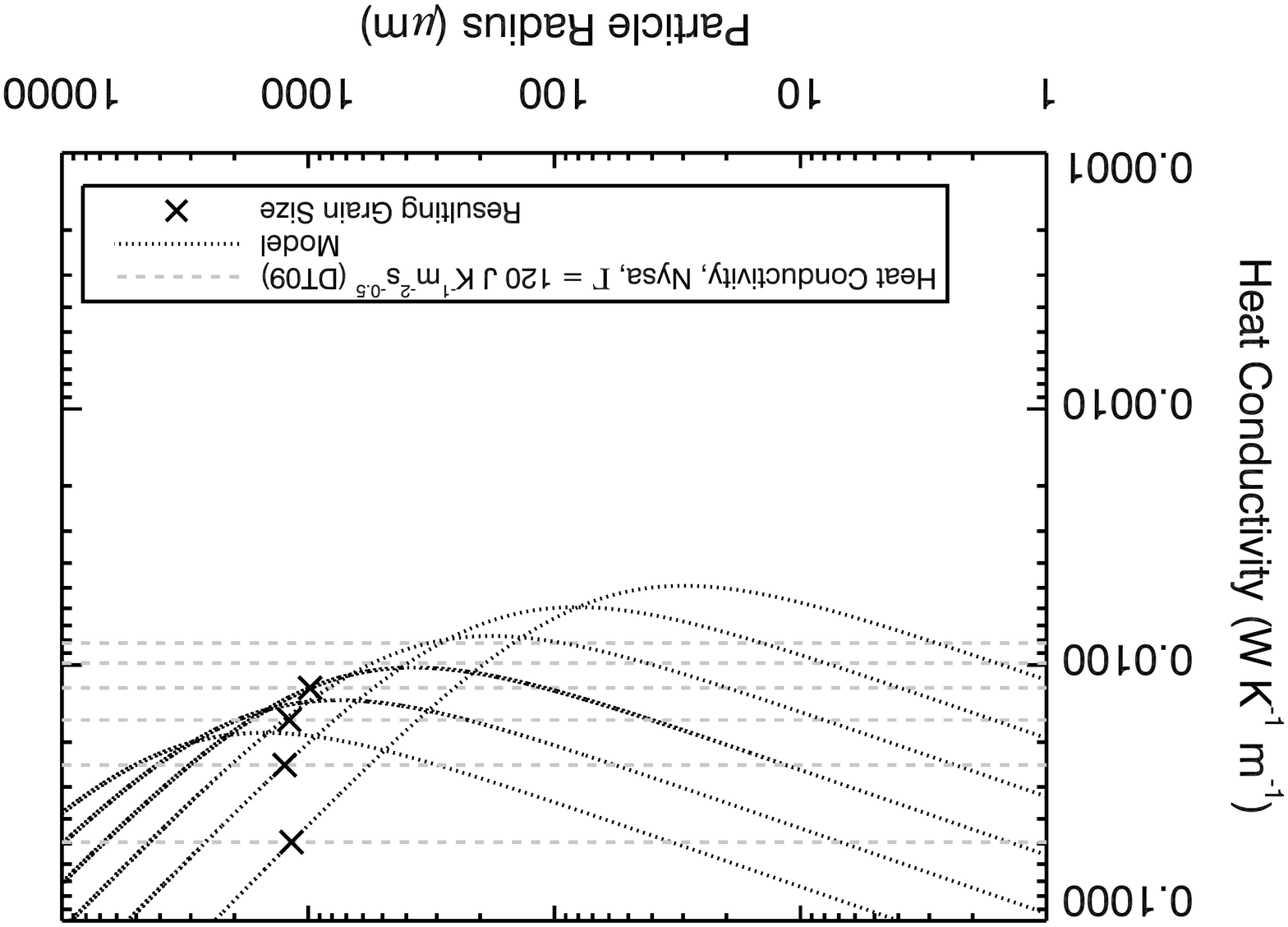}
\put(23,66){\Large a)}
\end{overpic}
\begin{overpic}[angle=180,width=1\columnwidth]{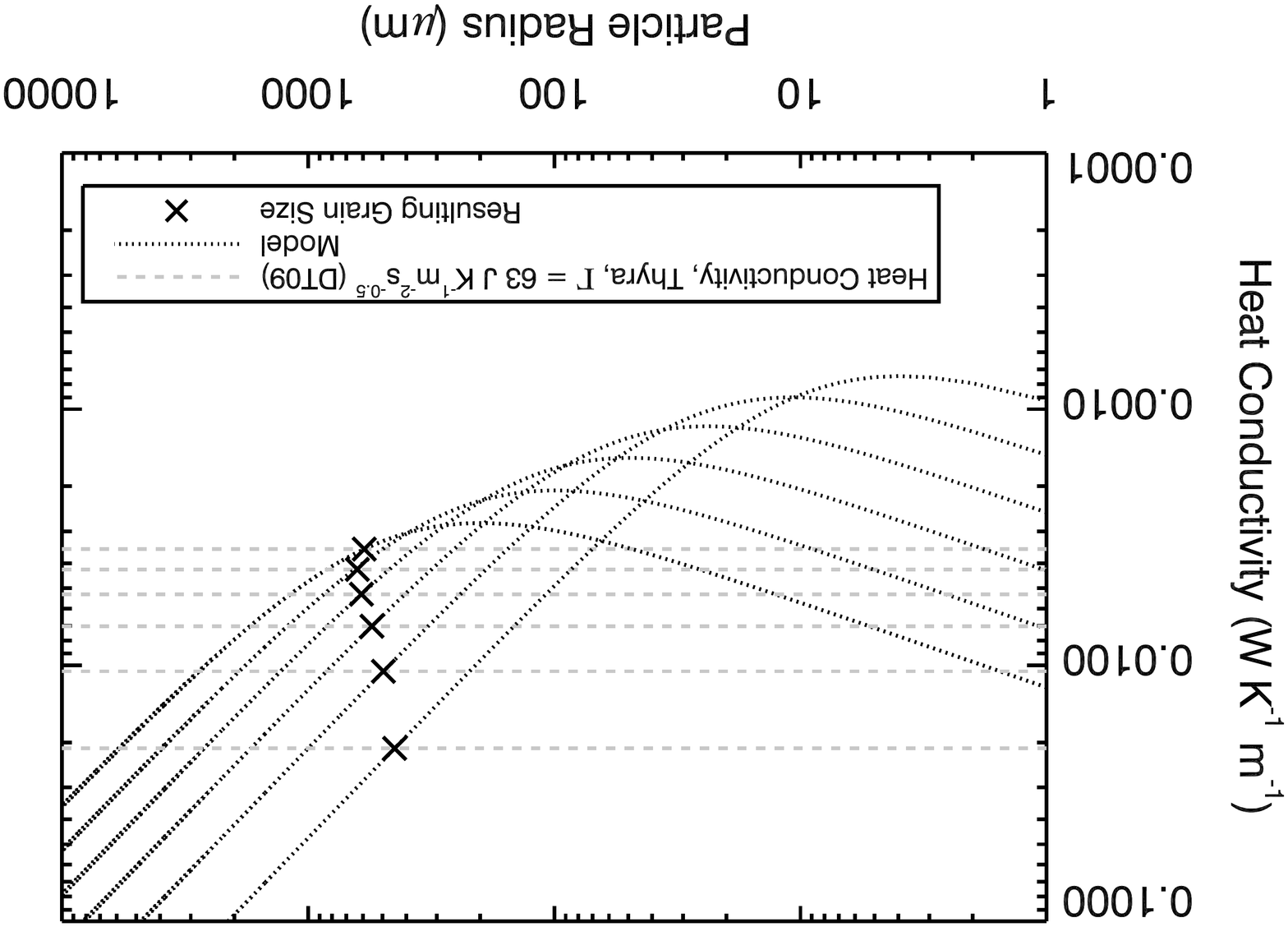}
\put(23,66){\Large b)}
\end{overpic}
\begin{overpic}[angle=180,width=1\columnwidth]{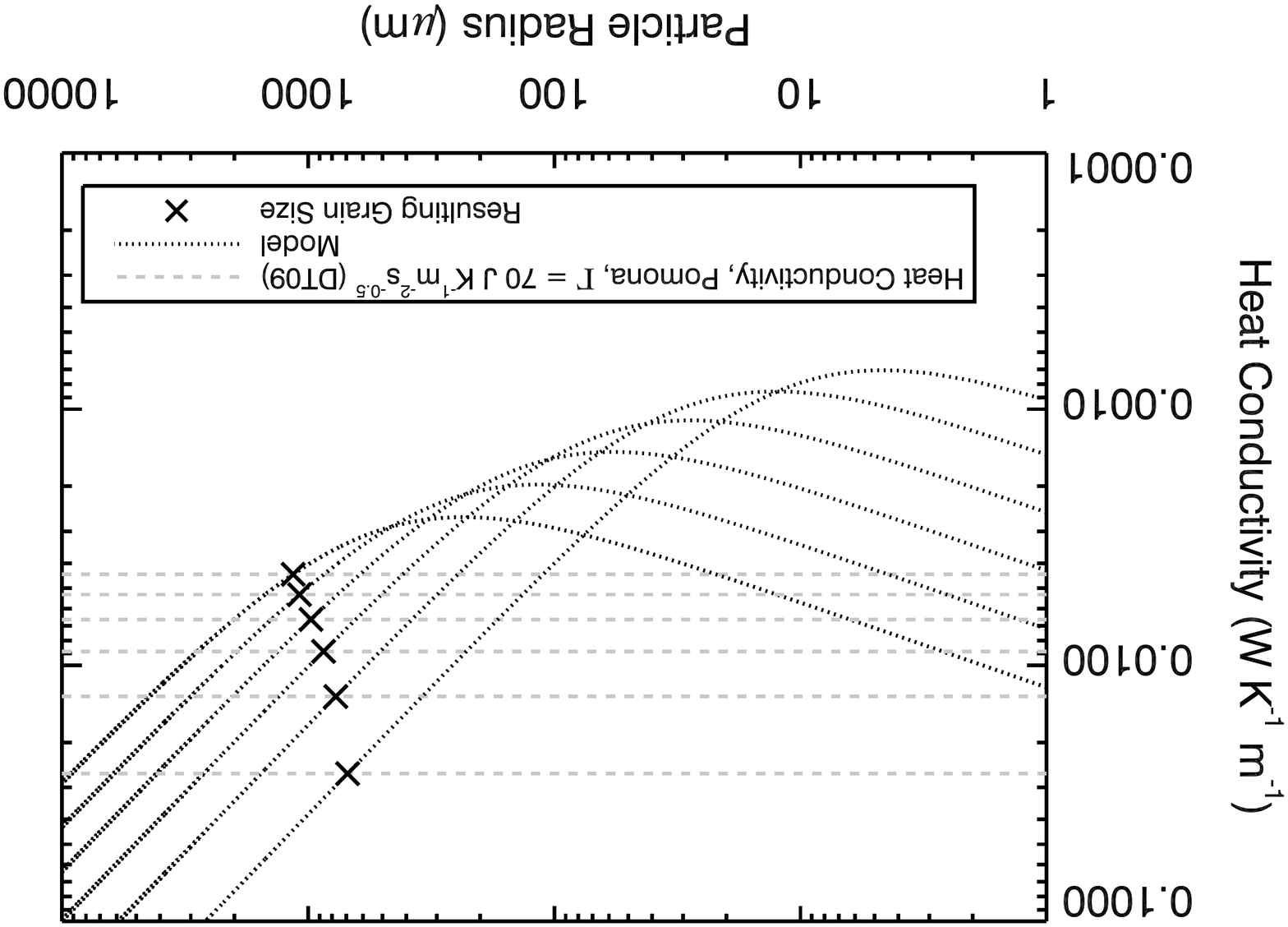}
\put(23,66){\Large c)}
\end{overpic}
\begin{overpic}[angle=180,width=1\columnwidth]{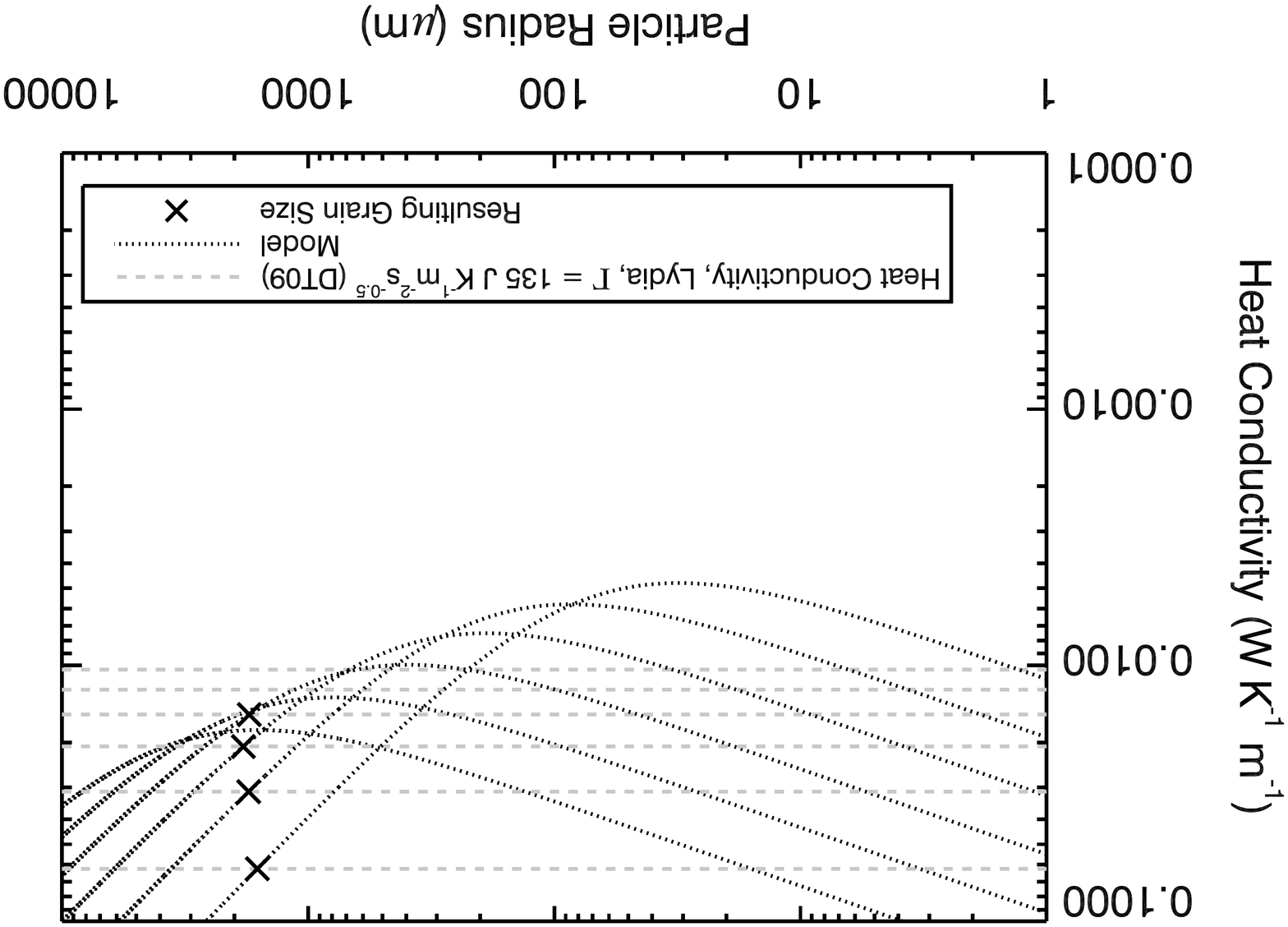}
\put(23,66){\Large d)}
\end{overpic}
\begin{overpic}[angle=180,width=1\columnwidth]{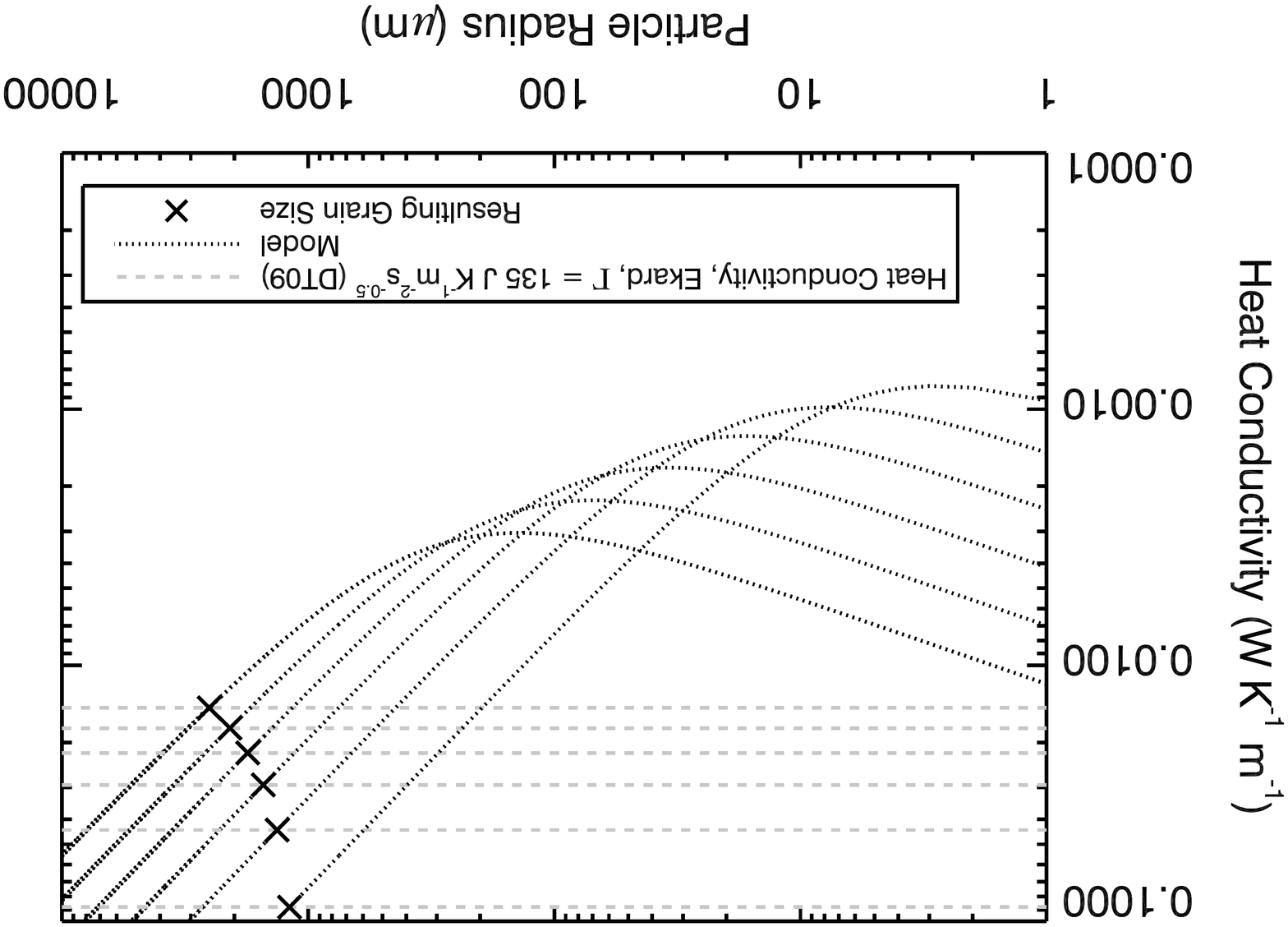}
\put(23,66){\Large e)}
\end{overpic}
\begin{overpic}[angle=180,width=1\columnwidth]{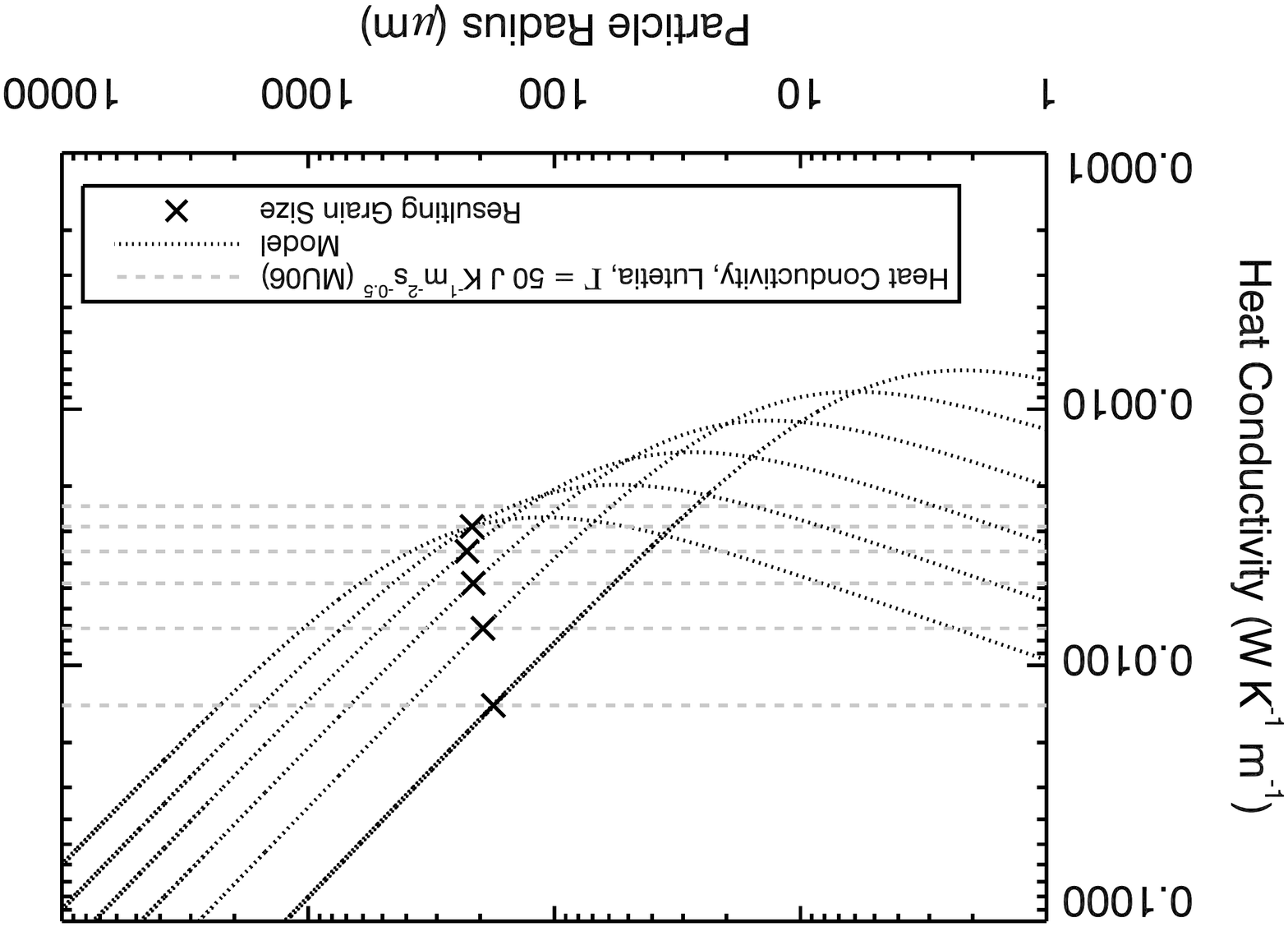}
\put(23,66){\Large f)}
\end{overpic}
\caption{Grain size estimation for the surface regolith of Nysa (a), Thyra (b), Pomona (c), Lydia (d), Ekard (e) and Lutetia (f).}
\label{Figcomp3}
\end{figure*}
\begin{figure*}[p!]
\centering
\begin{overpic}[angle=180,width=1\columnwidth]{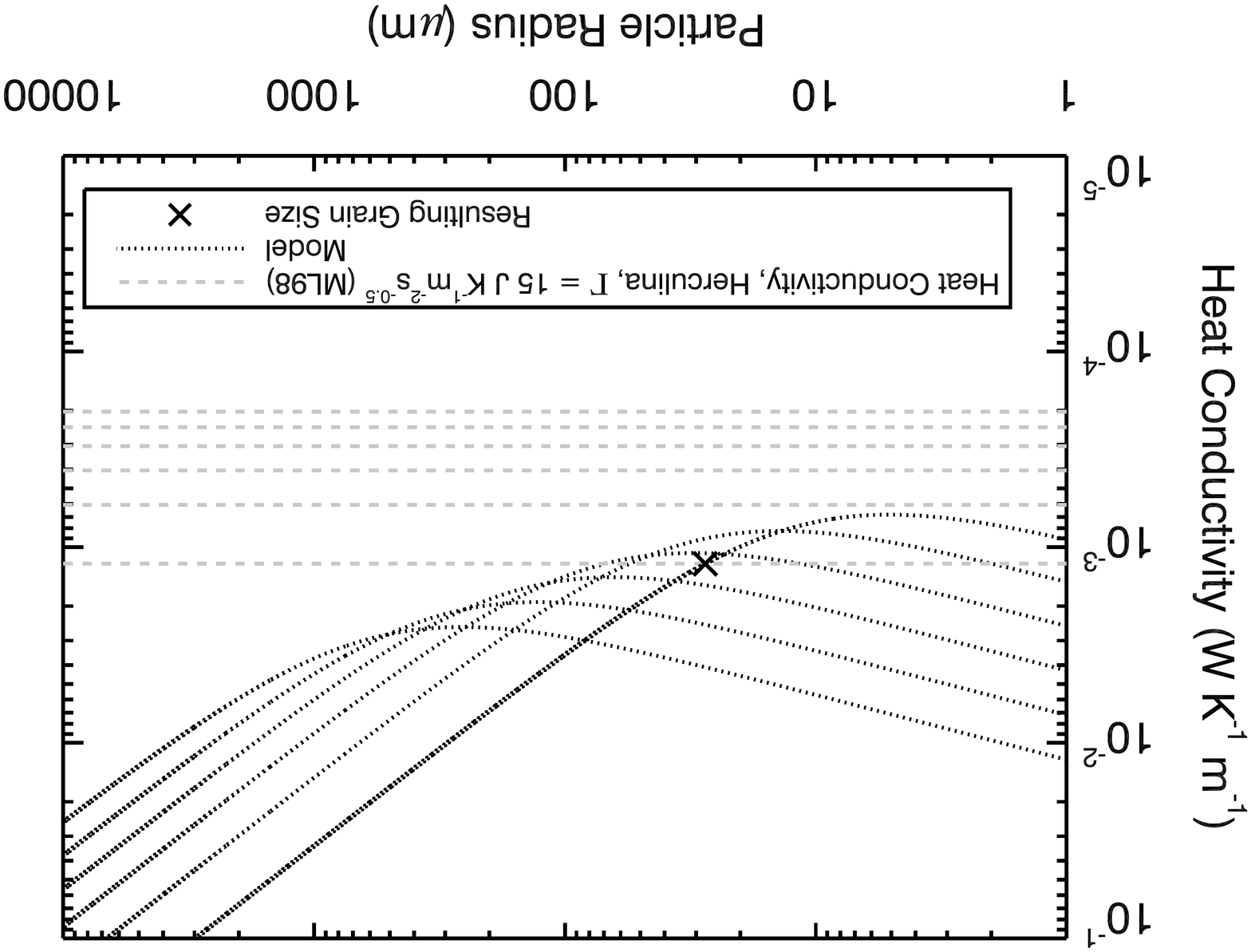}
\put(23,66){\Large a)}
\end{overpic}
\begin{overpic}[angle=180,width=1\columnwidth]{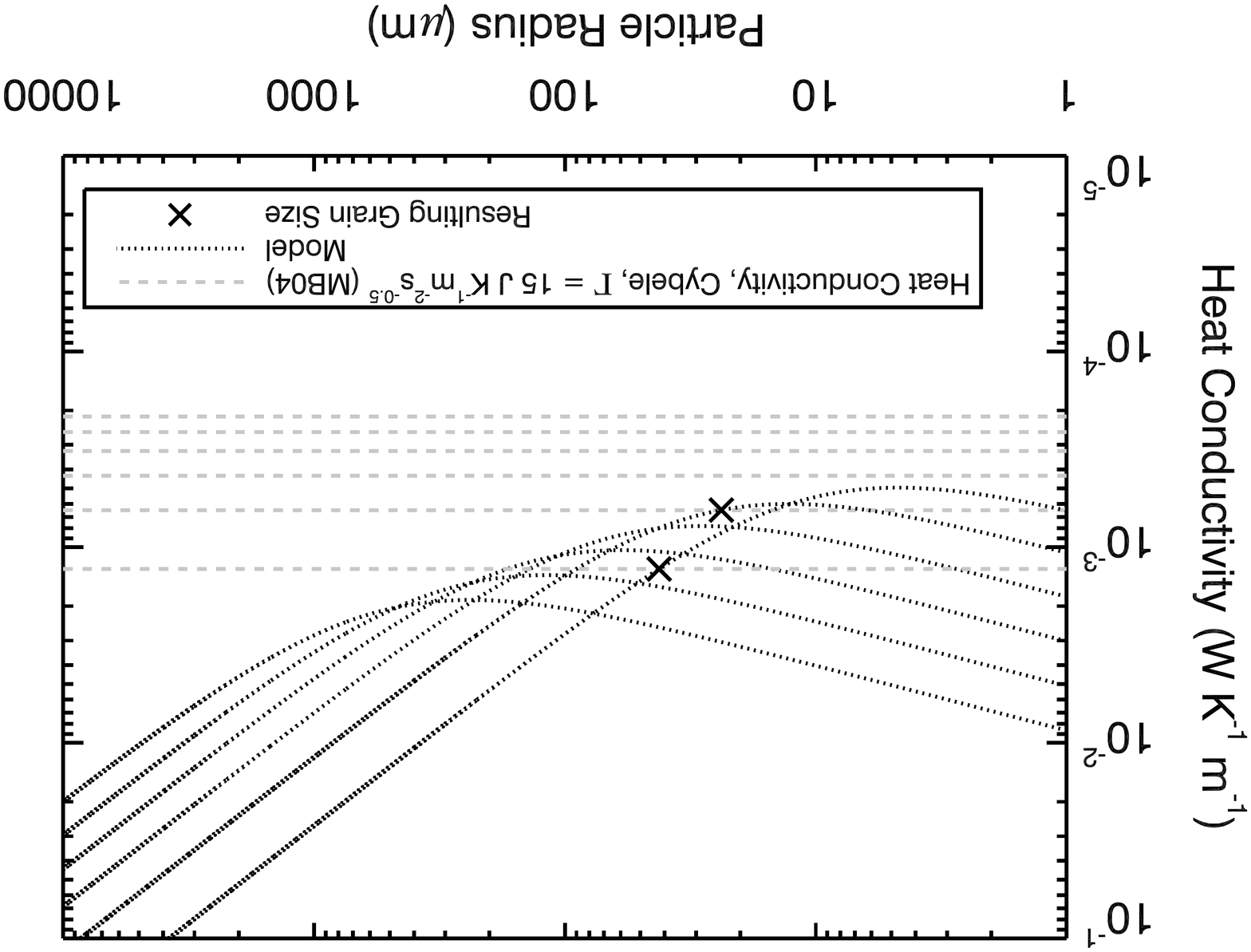}
\put(23,66){\Large b)}
\end{overpic}
\begin{overpic}[angle=180,width=1\columnwidth]{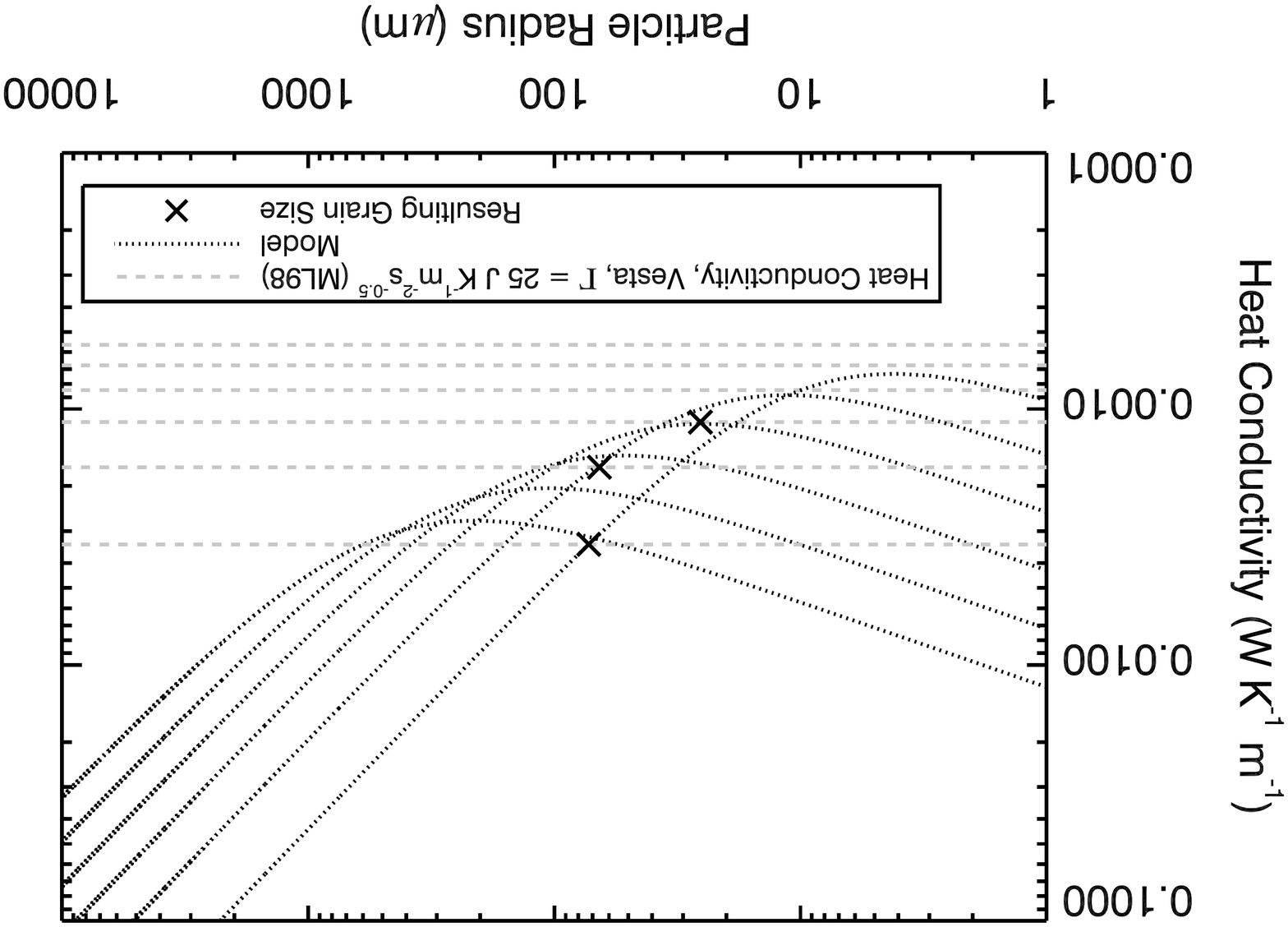}
\put(23,66){\Large c)}
\end{overpic}
\begin{overpic}[angle=180,width=1\columnwidth]{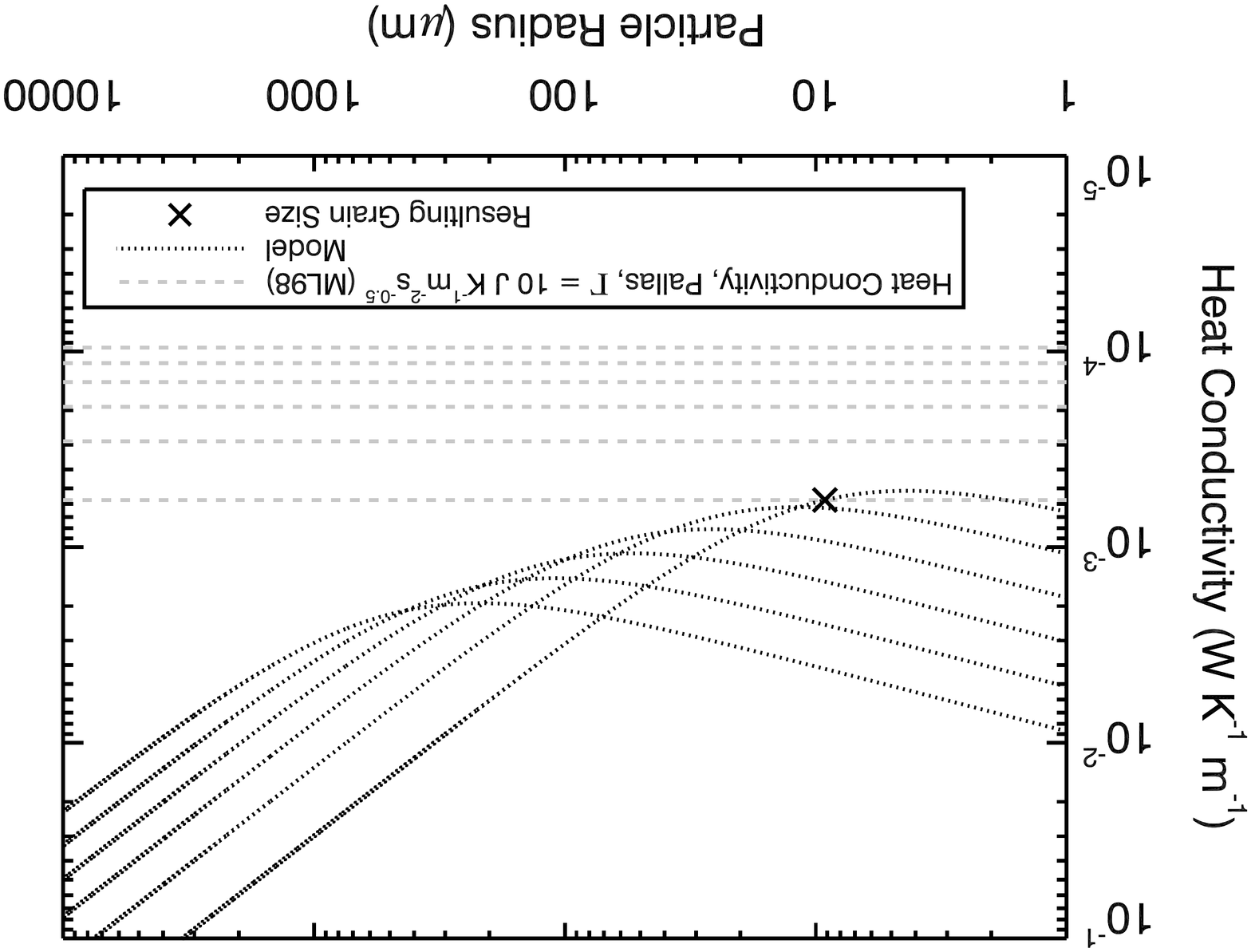}
\put(23,66){\Large d)}
\end{overpic}
\begin{overpic}[angle=180,width=1\columnwidth]{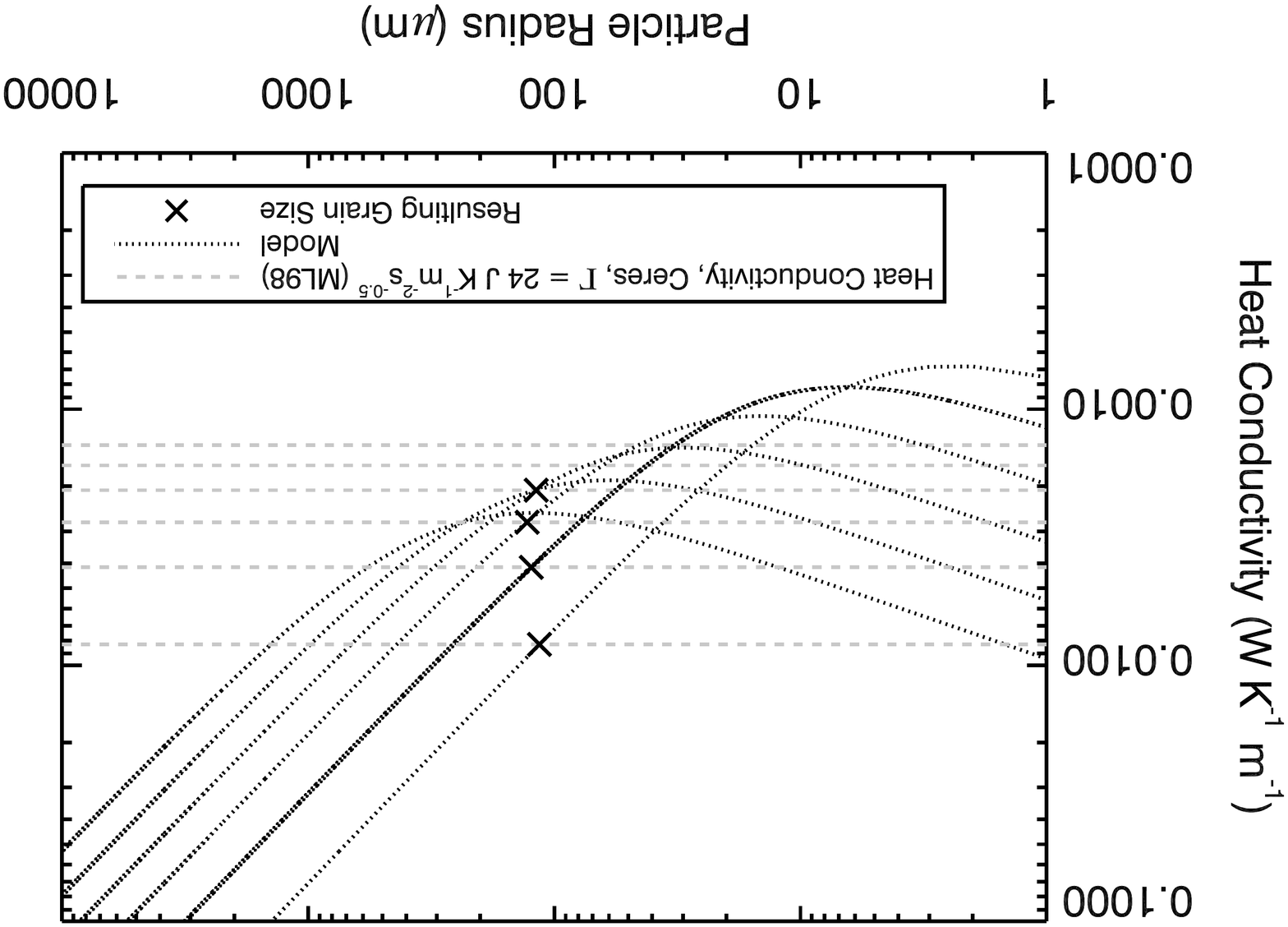}
\put(23,66){\Large e)}
\end{overpic}
\begin{overpic}[angle=180,width=1\columnwidth]{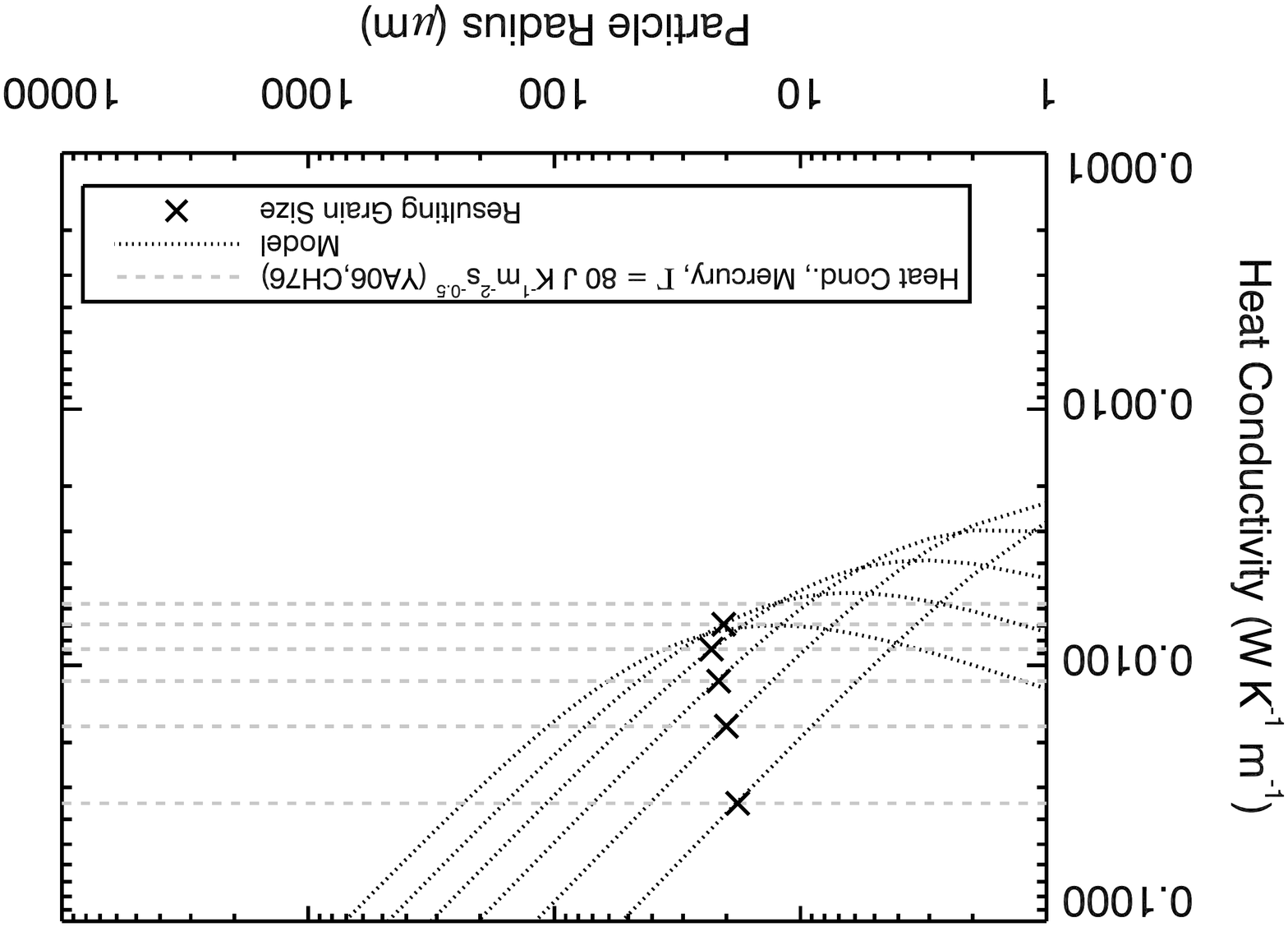}
\put(23,66){\Large f)}
\end{overpic}
\caption{Grain size estimation for the surface regolith of Herculina (a), Cybele (b), Vesta (c), Pallas (d), Ceres (e) and Mercury (f).}
\label{Figcomp4}
\end{figure*}

Figs. \ref{Figcomp1}-\ref{Figcomp4} show the grain size determination for all other objects listed in Tab. \ref{table_1}. The surface temperature of the objects was taken from the literature or was calculated (see Sect. \ref{Important properties of the analyzed objects}) if no direct temperature measurement was available.
\par
For the Moon and Mercury, the thermal inertia measurements were performed for a wide range of temperatures (see Tab. \ref{table_1}). For the Moon, we calculated the mean particle size using a surface temperature of $T \, = \, 295\, \mathrm{K}$. For Mercury, a surface temperature of $T\, = \, 700\, \mathrm{K}$ was used for the estimation of the mean grain size. In these two cases, the temperature variation causes an additional uncertainty of the grain size estimation. Both errors stemming from the temperature range are on the same order of magnitude. Hence, two error bars are later shown in Fig. \ref{Fig8} for Mercury, one for the uncertainty of the grain size estimation due to the temperature variation and the other for the uncertainty due the error of the thermal inertia measurement. For the Moon, unfortunately no error discussion of the thermal inertia estimation is available \citep{Wesselink1948}. Thus, the uncertainty of the lunar regolith grain size estimation is determined by the temperature range only.
\par
In some cases, the grain size estimation was not possible for the entire error interval of the measured thermal inertiae. For example, the thermal inertia of the asteroid Dodona was estimated to be $83 \pm 68 \, \mathrm{J \, m^{-2} \, K^{-1} \, s^{-0.5}}$ \citep{Delbo2009}, and we only found solutions for thermal inertiae ranging from $36 \,  \mathrm{J \, m^{-2} \, K^{-1} \, s^{-0.5}}$ to $151 \,  \mathrm{J \, m^{-2} \, K^{-1} \, s^{-0.5}}$. For values below $36 \,  \mathrm{J \, m^{-2} \, K^{-1} \, s^{-0.5}}$, the required values for the density of the surface regolith are too high. Tab. \ref{table_4} summarizes the cases, where only part of the measured thermal inertia range could be used to estimate the grain size of the regolith. The range of thermal inertiae used in our calculations is represented by $\Gamma'$ in Tab. \ref{table_4}.

\begin{table*}
\begin{center}
    \scriptsize
    \caption{Thermal-inertia ranges measured ($\Gamma$) and used in this work ($\Gamma'$) .}\vspace{1mm}
    \begin{tabular}[h!]{lr@{$\,$}lc@{$\,$}c}
    \hline
    Object & \multicolumn{3}{l}{$\Gamma$ [$\mathrm{J \, m^{-2} \, K^{-1} \, s^{-0.5}}$]} & $\ \ \ \ \ \ \Gamma'$ [$\mathrm{J \, m^{-2} \, K^{-1} \, s^{-0.5}}$]             \\\hline
    Dodona               &    83 & $^{+68}_{-68}$& \citep{Delbo2009}     &  $38 \, - \, 151$   \\
    Herculina            &    15 & $^{+8}_{-8  }$& \citep{Mueller1998}   &  $12 \, - \, 23$    \\
    Cybele               &    15 & $^{+8}_{-8  }$& \citep{Mueller2004}   &  $9 \, - \, 23$    \\
    Pallas               &    10 & $^{+5}_{-5}  $& \citep{Mueller1998}   &  $10 \, - \, 15$    \\
    \hline
    \multicolumn{4}{l}{$\Gamma$: measured thermal inertia.}\\[-0.15cm]
    \multicolumn{4}{l}{$\Gamma'$: used thermal inertia.}\\
        \end{tabular}
     \label{table_4}
\end{center}
\end{table*}

\section{Functional dependence of the mean regolith grain size on the gravitational acceleration of the asteroid}\label{Functional dependence of the mean regolith grain size on the size of the asteroid}
\begin{figure*}
\begin{overpic}[angle=0,width=1.0\textwidth]{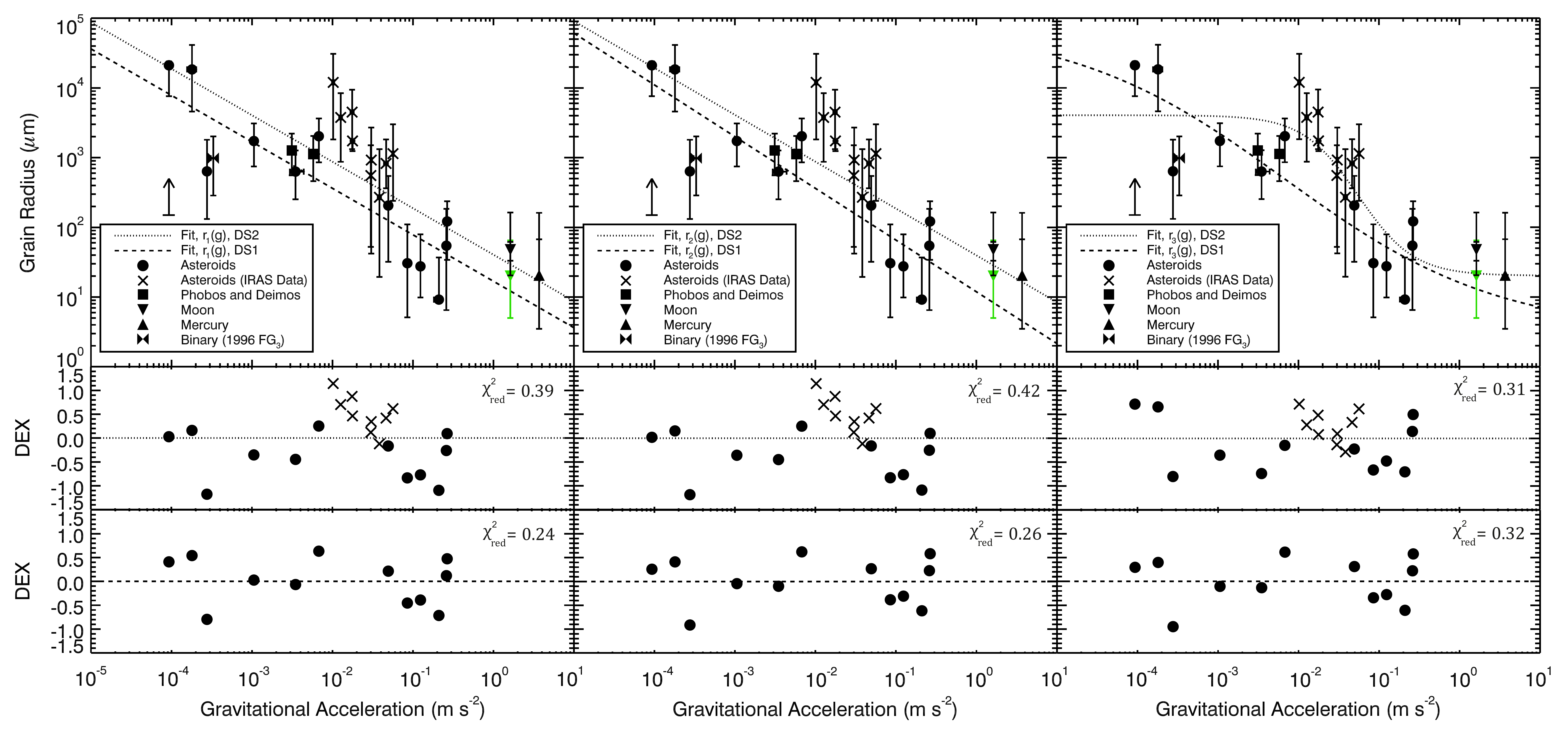}
\put(7.5,42.7){\Large a)}
\put(38,42.7){\Large b)}
\put(68.5,42.7){\Large c)}
\end{overpic}
\caption{Dependence of the regolith grain size on the gravitational acceleration of the asteroid. The derived grain sizes for the regolith of the asteroids in the sample of \citet{Delbo2007} and \citet{Delbo2009} as well as for the asteroids 1998 WT$_{24}$, 1999 JU$_{3}$ (the possible target of the Hyabusa 2 mission), 1996 FG$_3$ (the possible target of the Marco Polo-R mission), (2867) Steins, the Martian moons Phobos and Deimos, and the planet Mercury are shown as a function of the gravitational acceleration of the planetary body. Three different fit functions (dotted and dashed curves), $r_1(g)$ (a), $r_2(g)$ (b) and $r_3(g)$ (c), were fit to two different data sets (see Tab. \ref{table_5}). The first data set (DS1) consists of all asteroids without the IRAS asteroids (filled circles). The second data set (DS2) consists of all asteroids including the IRAS asteroids (filled circles plus crosses). The binary asteroid 1996 FG$_3$ (bowtie), the Martian moons Phobos and Deimos (filled squares), the Moon (downward triangle), and Mercury (upward triangle) were excluded from the fits. The green colored data denotes the measured size distribution of the lunar regolith samples \citep{McKay2009} and the arrow denotes the lower limit of the grain radius on the surface of Itokawa \citep{Yano2006,Kitazato2008}. Additionally, the residuals of the data to the fit functions are shown in logarithmic units for DS1 (lower row) and DS2 (upper row).}
\label{Fig8}
\end{figure*}

We applied our model to a number of asteroids for which thermal inertia measurements were available (see Sect. \ref{Determination of the regolith particle size}). Delbo and colleagues \citep{Delbo2007,Delbo2009} compiled such results and found that an inverse correlation between the thermal inertia and the diameter of the asteroid exists. We used these data and augmented them by thermal inertia measurements of the asteroids 1998 WT$_{24}$, 1999 JU$_{3}$ (the possible target of the Hayabusa 2 mission), 1996 FG$_3$ (the possible target of the Marco Polo-R mission) and (2867) Steins as well as of the Martian moons Phobos and Deimos and the planet Mercury. We sorted these objects into classes, with each class having its own set of material parameters, i.e. a temperature dependent heat conductivity and material density (see Tabs. \ref{Table_2} and \ref{table_1}). As done for the Moon, we varied the unknown packing density of the regolith from $\phi = 0.1$ to $\phi = 0.6$ and thus derived six different heat conductivity values for each body. We then applied our model (Eqs. \ref{eq1}-\ref{eq3}) and searched for intersections between the model and the respective heat conductivities, which yielded the grain size of the regolith (see Figs. \ref{Fig2} and \ref{Figcomp1}-\ref{Figcomp4}). We calculated the mean grain size by arithmetically averaging the (up to six) solutions. In the rare cases of double intersections, we chose the larger of the two possible values of the grain size. In Fig. \ref{Fig8}, we show the results of our analysis. As for the thermal inertiae, the derived mean particle size of the regolith follows an anti-correlation with the diameter and, thus, with the gravitational acceleration of the body (see Tab. \ref{table_1}). Small objects, like Asteroid (25143) Itokawa with a diameter of only 320 m and a gravitational acceleration of only $\sim 9.3 \times 10^{-5} \, \mathrm{m \, s^{-2}}$, possess a much coarser regolith, with typical particle sizes in the millimeter to centimeter regime, whereas those bodies exceeding 100 km in size ($g \gtrsim 5 \times 10^{-2} \, \mathrm{m \, s^{-2}}$) possess a much finer regolith, with grain sizes between 10 $\rm \mu m$ and 100 $\rm \mu m$. The uncertainties in the determination of the particle size (error bars in Fig. \ref{Fig8}) are dominated by the errors of the thermal inertia measurements, which are typically 50\% of the mean value (see Tab. \ref{table_1}).
\par
The decrease in grain size can qualitatively be understood as a result of the collision history of the asteroids. As mentioned above, previous studies have shown that hyper-velocity impacts lead to a size discrimination of the regolith, which consists of those impact fragments whose velocities did not exceed the escape speed of the parent body \citep{Chapman1976,Housen1982,Chapman2004}. As naturally the smaller fragments gain the highest velocities in impacts \citep{Fujiwara1980,Nakamura1991,Nakamura1993,Nakamura1994,Vickery1986,Vickery1987} and as the asteroids were roughly subject to the same impactor sizes and velocities, the expectation that small bodies are covered by the coarser fragments, whereas large bodies were able to hold on to most of the size distribution, seems to be justified (it should, however, be borne in mind that electrostatic effects might also play a role for the release or capture of surface dust \citep{Lee1996}). Typical impact generated size distribution functions have the tendency to be dominated in number by the smallest fragments \citep{Nakamura1994} so that the mean grain size responsible for the thermal conductivity of the regolith is expected to be close to the smaller limit.
\par
To analyze the data, we fitted the logarithmic data with the following functions,
\begin{align}
&\mathrm{log} \left[\frac{r_1(g)}{\mathrm{1\,\mu m}}\right] \, = \, b_1 - \, \frac{2}{3} \, \mathrm{log} \left[\frac{g}{\mathrm{1\, m\,s^{-2}}}\right] \, \mathrm{,} \label{new_eq_fit} \\[2mm]
&\mathrm{log} \left[\frac{r_2(g)}{\mathrm{1\,\mu m}}\right] \, = \, b_1 - \, b_2 \, \mathrm{log} \left[\frac{g}{\mathrm{1\,m\,s^{-2}}}\right]
\label{new_eq_fita}
\end{align}
and
\begin{align}
\mathrm{log} \left[\frac{r_3(g)}{\mathrm{1\,\mu m}}\right] \, = \, \frac{b_1}{1 \, + \, \mathrm{exp}\left[\, b_2 \, ( \, \mathrm{log} \left[\frac{g}{\mathrm{1\,m\,s^{-2}}}\right]  \, - \, b_3 \, )\right]} \, + \, b_4
\label{new_eq_fitb}
\end{align}
(Figs. \ref{Fig8}a-c). Tab. \ref{table_5} summarizes the used fit functions, the values of the fit parameters, and the resulting values of the reduced chi square, $\chi^2_{\rm red}$. The power law fits $r_1 \propto g^{-2/3}$ and $r_2 \propto g^{-b_2}$ are motivated by the fact that laboratory experiments \citep{Fujiwara1980,Nakamura1991,Nakamura1993,Nakamura1994} and studies of the crater structures on the Moon \citep{Vickery1986,Vickery1987} have shown that hyper-velocity impacts accelerate small fragments to higher velocities than large ones. Fig. 5 in \citet{Nakamura1994} shows a comparison of the laboratory data and the results of the crater structure analysis. A simple interpolation of the data on the upper left and lower right of Fig. 5 in \citet{Nakamura1994} by a power law function yields a dependency between the velocity $v_{frag}$ and the size $d_{frag}$ of the ejected fragments of $v_{frag} \, \sim \, d_{frag}^{-3/2}$. Note that Fig. 5 in \citet{Nakamura1994} visualizes the velocity of the fragments on the vertical axis and the size of the fragments on the horizontal axis. However, our Fig. \ref{Fig8} shows the size of the regolith on the vertical axis and the gravitational acceleration, which is proportional to the escape speed of the body, on the horizontal axis. We have inverted the estimated exponent in order to apply the power law function to our data. Hence, the grain size of the asteroidal regolith particles should follow a power law function of the surface acceleration with an exponent of $\sim - 2/3$. The smoothed step function $r_3(g)$ was chosen because of the appearance of the data, which suggest a rather steep decrease of the regolith grain size for bodies of $\sim 100$ km in size ($g = \sim 5 \times 10^{-2} \, \mathrm{m \, s^{-2}}$), close to the value above which asteroids are thought to be primordial \citep{Bottke2005}.
\par
All fit functions were applied to two different data sets (DS1 and DS2). The first data set (DS1; filled circles in Fig. \ref{Fig8}) consists of all asteroids without the asteroids analyzed using the IRAS data (short: IRAS asteroids). The second data set (DS2; filled circles plus crosses in Fig. \ref{Fig8}) consist of all asteroids including the IRAS asteroids. The binary asteroid 1996 FG$_3$ (bowtie), the Martian moons Phobos and Deimos (filled squares), the Moon (downward triangle) and Mercury (upward triangle) were excluded from the fits because the surface regolith on these objects was probably influenced by the presence of the close-by asteroid (1996 FG$_3$) or planet (Mars or Earth), or by the location deep within the solar gravitational potential (Mercury).
\par
We found that fixing the exponent of the power law to $- 2/3$ results in a slightly better fit than using the exponent as a free parameter (see Tab. \ref{table_5}). On top of that, both power laws are identical within the errors of the fit parameters. As was already addressed in \citet{Delbo2009}, the thermal inertiae of the asteroids observed by IRAS seem to be offset and to possess a steeper slope than the rest of the sample. Excluding this sample from the above analysis yields a better fit of the power law functions to the data.
\par
However, the smoothed step function yields a better fit to the data if the asteroids observed by IRAS are taken into account (see Tab. \ref{table_5}). Excluding the IRAS asteroids decreases the quality of the fit in case of the smoothed step function. The residuals of the smoothed step function including the IRAS asteroids are almost identical to the residuals of the two power law functions without the IRAS data. The smoothed step function was chosen due to the appearance of the data and has no physical explanation yet. However, electrostatic effects close to the surface of the small asteroids have been mentioned to influence the escape of ejecta \citep{Lee1996}. The smoothed step function obviously shows a separation of the objects into two different classes. Small objects with diameters less than $\sim  20 \, \mathrm{km}$ ($g \lesssim 1 \times 10^{-2} \, \mathrm{m \, s^{-2}}$) are covered by relatively coarse regolith grains with typical particle sizes in the millimeter to centimeter regime, whereas large objects with diameters in excess of $\sim  80 \, \mathrm{km}$ ($g \gtrsim 6 \times 10^{-2} \, \mathrm{m \, s^{-2}}$) are covered by very small regolith particles with grain sizes between 10 $\rm \mu m$ and 100 $\rm \mu m$.
\begin{table*}
\begin{center}
    \scriptsize
    \caption{Fit functions and resulting fit parameters used to describe the anti-correlation between regolith particle size and gravitational acceleration of the asteroid.}\vspace{1mm}
    \begin{tabular}[h!]{lcccccc}
    \hline
    Fit Function &Data Set& $b_1$ & $b_2$ & $b_3$ & $b_4$ & $\chi^2_{\rm red}$          \\\hline
    Eq. \ref{new_eq_fit} & DS1 & $1.23 \, \pm \, 0.14$ & - & - & - & $0.24$  \\
    Eq. \ref{new_eq_fit} & DS2 & $1.61 \, \pm \, 0.14$ & - & - & - & $0.39$  \\
    Eq. \ref{new_eq_fita} & DS1 & $1.08 \, \pm \, 0.27$& $0.74 \, \pm \, 0.12$ & - & - & $0.26$  \\
    Eq. \ref{new_eq_fita} & DS2 & $1.60 \, \pm \, 0.30$& $0.67 \, \pm \, 0.14$ & - & - & $0.42$   \\
    Eq. \ref{new_eq_fitb} & DS1 & $4.43 \, \pm \, 0.00^\dag $ & $0.75 \, \pm \, 0.00^\dag $ & $-2.17 \, \pm \, 0.00^\dag $ & $0.48 \, \pm \, 0.00^\dag $ & $0.32$  \\
    Eq. \ref{new_eq_fitb}& DS2 & $2.30 \, \pm \, 0.00^\ddag $ & $2.90 \, \pm \, 0.00^\ddag $ & $-1.26 \, \pm \, 0.00^\ddag $ & $1.31 \, \pm \, 0.00^\ddag $ & $0.31$  \\
    \hline
    \multicolumn{7}{l}{$\chi^2_{\rm red}$: reduced chi square.}\\[-0.15cm]
    \multicolumn{7}{l}{DS1: data set 1; all asteroids excluding the IRAS asteroids.}\\[-0.15cm]
    \multicolumn{7}{l}{DS2: data set 2; all asteroids including the IRAS asteroids.}\\[-0.15cm]
    \multicolumn{7}{l}{$^\dag$: the uncertainties of the fit parameters are: $\Delta b_1 = 1.00 \times 10^{-4}$, $\Delta b_2 = 4.94 \times 10^{-4}$, $\Delta b_3 =  7.63 \times 10^{-4}$ and $\Delta b_4 = 5.20 \times 10^{-4}$.}\\[-0.15cm]
    \multicolumn{7}{l}{$^\ddag$: the uncertainties of the fit parameters are: $\Delta b_1 = 5.27 \times 10^{-5}$, $\Delta b_2 = 3.15 \times 10^{-4}$, $\Delta b_3 =  3.56 \times 10^{-5}$ and $\Delta b_4 = 3.82 \times 10^{-5}$.}\\[-0.15cm]
        \end{tabular}
     \label{table_5}
\end{center}
\end{table*}

\section{Influence of the network heat conductivity on the grain size estimation}\label{Influence of the network heat conductivity on the grain size estimation}
As can be seen above, small asteroids ($\lesssim 20 \, \mathrm{km}$) are covered by relatively coarse regolith (in the millimeter to centimeter regime, see Sect. \ref{Functional dependence of the mean regolith grain size on the size of the asteroid}). In the case of coarse regolith, the total heat conductivity of the regolith is dominated by radiative heat transport. Here, we show that the regolith grain size of small asteroids ($\lesssim 50 \, \mathrm{km}$) can be easily estimated by only calculating the radiative heat transport of the regolith, $\lambda \approx \lambda_{\rm rad} = \, 8 \, \sigma \, \epsilon \, T^3 \, \Lambda(r,\phi)$ (see Eq. \ref{eq1}). This approximation provides grain sizes with an accuracy of approximately a factor of 2.
\par
\begin{figure}[b!]
\centering
\begin{overpic}[angle=180,width=1\columnwidth]{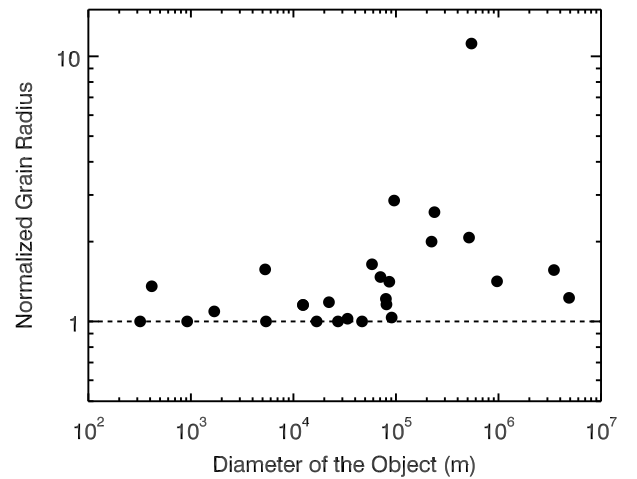}
\end{overpic}
\caption{This figure shows the influence of the solid network heat conductivity on the grain size estimation by comparing the normalized grain radius $r' / r$ with the diameter of the object. The grain radius $r'$ was estimated under the assumption that heat is only transported by radiation, i.e. $\lambda_{\rm solid}(T) \, H(r,T,\phi) \, \ll \, 8 \, \sigma \, \epsilon \, T^3 \, \Lambda(r,\phi)$. The deviation of the data from the dashed line shows the importance of the heat conductivity of the solid network of regolith on the grain size estimation.}
\label{Fig7}
\end{figure}

Fig. \ref{Fig7} shows the normalized grain radius $r' / r$ as a function of the diameter of the object. Here, $r'$ is the derived grain size of the surface regolith particles under the assumption that the heat transport is due to radiation, i.e. $\lambda_{\rm solid}(T) \, H(r,T,\phi) \, \ll \, 8 \, \sigma \, \epsilon \, T^3 \, \Lambda(r,\phi)$. If the heat conductivity was solely determined by radiative transport, all data would be located on the dashed line, $r'/r=1$. The deviation of the data from the dashed line shows the influence of the heat conductivity of the solid network of regolith on the grain size estimation. For small objects ($\lesssim 50 \, \mathrm{km}$), the deviation of the relative grain radius to the dashed line is relatively small (less than a factor of 2). This demonstrates that for small objects ($\lesssim 50 \, \mathrm{km}$), the total heat conductivity of the regolith is generally dominated by radiative heat transport. In this case the estimated grain sizes are independent of the specific material properties, as long as the condition $\epsilon \approx 1$ still holds (see Eqs. \ref{eq1} and \ref{eq6}). For larger objects ($\gtrsim 50 \, \mathrm{km}$), the heat conductivity of the solid network of regolith and, thus, the material properties are not negligible.

\section{Conclusion and discussion}\label{Discussion}
In this work, we presented a new method to determine the grain size of regolith particles by using remote measurements only. We utilized measurements of the thermal inertia performed for the regolith of various objects in the Solar System (see Tab. \ref{table_1}). With the knowledge of the thermal inertia, the surface-material properties and assumptions about the packing density of the regolith particles, the heat conductivity of the surface regolith was derived. The grain size of the planetary regolith was then determined from a comparison of the derived heat conductivity with a modeled heat conductivity of granular materials in vacuum.
\par
\begin{figure}[b!]
\centering
\begin{overpic}[angle=180,width=1.0\columnwidth]{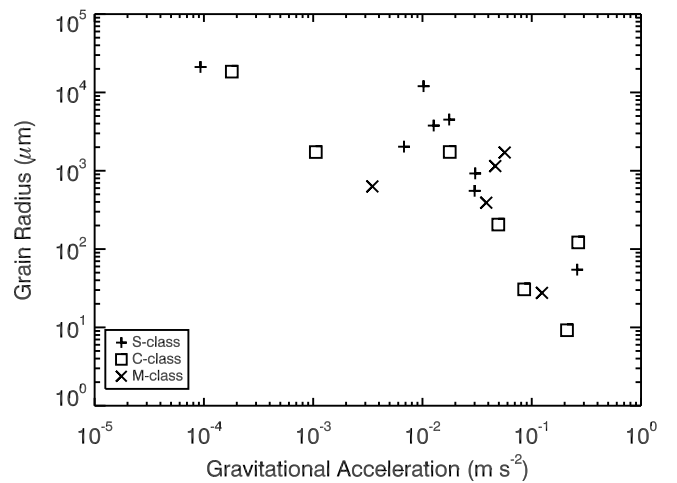}
\end{overpic}
\caption{Comparison of the derived grain sizes for the different spectral classes of the asteroids: S-class (pluses), C-class (squares) and M-class (crosses).}
\label{Fig10}
\end{figure}

We determined the grain size of planetary regolith for a large number of asteroids, the Moon, the Martian moons and Mercury with diameters between 0.3 km and 4,880 km (see Tab. \ref{table_1}) and find an anti-correlation between the regolith grain size and the gravitational acceleration of the planetary body (see Fig. \ref{Fig8}). The anti-correlation between grain size and gravitational acceleration is independent of the spectral class of the asteroids (see Fig. \ref{Fig10}). This outcome supports the idea that planetary regolith is formed by hyper-velocity impacts, which have ground down the  initially rocky material to ever finer particle sizes \citep{Chapman1976,Housen1982,Chapman2004}.
\par
However, the determined grain sizes also show a separation of the objects into two different regimes. Small objects with diameters less than $\sim 20 \, \mathrm{km}$ ($g \lesssim 1 \times  10^{-2} \, \mathrm{m \, s^{-2}}$) are covered by relatively coarse regolith grains with typical particle sizes in the millimeter to centimeter regime and large objects with diameters bigger than $\sim 80 \, \mathrm{km}$ ($g \gtrsim 6 \times 10^{-2} \, \mathrm{m \, s^{-2}}$) possess very small regolith particles with grain sizes between 10 $\rm \mu m$ and 100 $\rm \mu m$. A previous estimation of the grain size of asteroidal surface regolith derived from polarization measurements \citep[grain size: $\sim 100 \, \mathrm{\mu m}$,][]{Dollfus1979} is in agreement with our result.
\par
Our method to determine the grain size of planetary regolith depends on the measured thermal inertia of the object, the surface temperature of the body during observation and the material properties of the regolith. Due to Eq. \ref{eq0}, the uncertainties of the thermal inertia measurements, which are typically $50 \, \%$ of the measured value, primarily determine the error of the grain size estimation (see Tab. \ref{table_1}). Typical temperature variations of the analyzed objects have only a minor influence on the grain size estimation (less than $\sim 10 \%$ of the estimated value). Only for the Moon and Mercury, the error of the grain size estimation due to the temperature variation becomes important (see Sect. \ref{Determination of the regolith particle size}). The used material properties only have an influence on the grain size estimation if the diameter of the object is larger than $\sim 50 \, \mathrm{km}$ (see Fig. \ref{Fig7} in Sect. \ref{Influence of the network heat conductivity on the grain size estimation}).
\par
In principle, a rough estimation (accuracy within a factor of two) of the grain size of the surface regolith of small objects can be performed by assuming radiative transport only (see Fig. \ref{Fig7}). For bigger objects and, thus, smaller regolith particles, the heat conductivity of the regolith network becomes more important so that the uncertainty of the grain-size estimation stemming from this simplification increases.
\par
We conclude that our method allows the direct determination of the grain size of asteroidal or planetary regolith from measurements of the thermal inertia of the parent body. The derived mean particle radii vary from millimeters to centimeters for km-sized objects to $10-100\rm ~\mu m$ for bodies exceeding $\sim 80$ km in diameter. This finding is important for the planning of future landing and/or sample-return missions as it allows narrowing down the physical state of the regolith by state-of-the-art remote observations. Smaller particle sizes in the regolith lead to a higher cohesion of the uppermost layers, which is, for small bodies, not determined by the local gravity field but dominated by inter-particle forces.
\par
On top of that, our method can be used to validate models of the formation and evolution of planetary regolith. Hypervelocity impacts cause the ejection of impact fragments with a characterisitc size distribution and size-velocity relation \citep{Fujiwara1980,Nakamura1991,Nakamura1993,Nakamura1994,Vickery1986,Vickery1987}. In combination with the impact geometry and the escape speed of the target object, one can thus derive a zero-age regolith size distribution. Space wheathering, granular transport effects, and other internal and external influences might alter this size distribution over time.

\subsection*{Acknowledgements}
We thank Patrick Michel, Akiko Nakamura and the ISSI team for fruitful discussions on regolith properties, the JPL small body database for providing the physical parameters of the asteroids and the reviewers for the improvements of the manuscript.

\bibliographystyle{model2-names}
\bibliography{bib}

\end{document}